\documentclass[printer]{aa} 
\usepackage{graphicx}
\usepackage{longtable}
\usepackage{natbib}
\bibpunct{(}{)}{;}{a}{}{,} 
\begin{document}
\title{A self-consistent  approach to  \\the hard and  soft states  of \object{4U 1705-44}
\thanks{Based  on observations obtained  with XMM-Newton, an ESA  science mission with  instruments and  contributions directly
funded by ESA Member States and NASA}}

\author{A. D'A\`i  \inst{1}
\and T. Di Salvo \inst{1}
\and D. Ballantyne \inst{2}      
\and R. Iaria \inst{1} 
\and N.R. Robba \inst{1} 
\and A. Papitto \inst{3,4} 
\and A. Riggio \inst{3,4} 
\and L. Burderi \inst{3} 
\and S. Piraino \inst{5,6} 
\and A. Santangelo \inst{5}
\and G. Matt \inst{7} 
\and M. Dov\v{c}iak \inst{8} 
\and V. Karas \inst{8} 
}
\institute{Dipartimento di Scienze Fisiche ed Astronomiche, Universit\`a di Palermo\\
\email{dai@fisica.unipa.it}
\and Center for Relativistic Astrophysics, School of Physics, Georgia Institute of Technology, Atlanta, GA 30332, USA
\and Universit\`a degli Studi di Cagliari, Dipartimento di Fisica, SP Monserrato-Sestu, KM 0.7, 09042 Monserrato, Italy
\and INAF -- Osservatorio Astronomico di Cagliari, Poggio dei Pini, Strada 54, 09012 Capoterra (CA), Italy
\and IAAT, University of Tubingen, Sand 1, 72076 Tubingen, Germany 
\and INAF – IASF di Palermo, via Ugo La Malfa 153, 90146 Palermo, Italy
\and Dipartimento di Fisica, Universit\`a degli Studi Roma Tre, Via della Vasca Navale 84, I-00146 Roma, Italy
\and Astronomical Institute, Academy of Sciences of the Czech Republic, Bocni II 1401a, CZ-14131 Prague, Czech Republic
}
\date{Received ... ; accepted ...}
\abstract{}{}{}{}{} 
  \abstract
 {High-resolution spectroscopy has recently revealed in many low-mass
 X-ray binaries hosting a neutron  star that the shape of the broad
 iron  line observed  in  the 6.4-6.97  keV  range is  consistently
 well-fitted by a relativistically smeared line profile.}
{The presence of other broad features, besides the iron line, together
  with a  high S/N of the  spectra offer the possibility  of testing a
  self-consistent  approach   to  the  overall   broadband  reflection
  spectrum and  evaluating the impact  of the reflection  component in
  the formation of the broadband X-ray spectra.}
{We analyzed two XMM-Newton observations of the bright atoll source \object{4U
  1705-44}, which  can be  considered a prototype  of the class  of the
  persistent NS  LMXBs showing both  hard and soft states.   The first
  observation was  performed when  the source was  in a hard  low flux
  state, the second  during a soft, high-flux state.  Both the spectra
  show  broad  iron  emission  lines.   We fit  the  spectra  using  a
  two-component model,  together with a  reflection model specifically
  suited to  the case of a  neutron star, where  the incident spectrum
  has a blackbody shape.}
{In  the   soft  state,  the   reflection  model,  convolved   with  a
  relativistic  smearing component,  consistently describes  the broad
  features  present in  the spectrum,  and  we find  a clear  relation
  between the temperature of the  incident flux and the temperature of
  the harder X-ray  component that we interpret as  the boundary layer
  emission.   In  this state  we  find  converging  evidence that  the
  boundary  layer outer  radius is  $\sim$  2 times  the neutron  star
  radius. In the low flux state,  we observe a change in the continuum
  shape of  the spectrum with respect  to the soft  state.  Still, the
  broad  local  emission  features  can  be  associated  with  a  disk
  reflecting  matter, but in  a lower  ionization state,  and possibly
  produced in an accretion disk  truncated at greater distance. } {Our
  analysis provides  strong evidence that the  reflection component in
  soft states of  LMXBs comes from to hard  X-ray thermal irradiation,
  which we identify with the  boundary layer emission, also present in
  the continuum model. In the hard  state, the broad iron line if also
  produced  by reflection,  and  the continuum  disk  emission can  be
  self-consistently accounted  if the disk  is truncated at  a greater
  distance than the soft state.}  
\titlerunning{Hard and soft states in 4U 1705-44}
\authorrunning{A. D'A\`i et al.}
\maketitle


\section{Introduction}

In  last few  years,  the Epic-pn  instrument  onboard the  XMM-Newton
satellite has  allowed deep investigation  of the nature of  the broad
emission  lines  observed  in  the  iron K$\alpha$  region  of  bright
accreting   neutron  star   (NS)  low-mass   X-ray   binaries  (LMXBs)
\citep{bhattacharyya07,   pandel08,   cackett08,   dai09,   papitto09,
  iaria09, disalvo09}.  The authors of these works  have focused their
attention on  the shape and  origin of the  broad iron line,  and they
agreed on the interpretation of the line broadness being the result of
to  special  and general  relativistic  effects  arising  in the  disk
reflecting  matter  at a  few  gravitational  radii  from the  compact
object.  This interpretation  is supported by theoretical expectations
and by  general agreement  between the fitting  model and  X-ray (1-10
keV) data.

However,  if this  interpretation is  correct, the  reflected spectrum
should encompass a variety  of other disk reflection features, because
there  are other  low-Z,  but  abundant, emitting  ion  metals in  the
low-energy band.  The good spectral resolution and the high S/N of the
spectra has effectively shown in  many cases a more complex pattern of
features,  besides the broad  iron line.   \citet{dai09} found  in the
spectrum  of the bright  Z-source GX  340+0 a  broad emission  line of
\ion{Ca}{xix}  and   an  absorbing   edge  of  highly   ionized  iron;
\citet{iaria09} found in  the spectrum of the Z-source  GX 349+2 three
broad  lines,  besides the  iron  line,  identified as  \ion{Ca}{xix},
\ion{Ar}{xviii} and  a blending  of L-shell transitions  of moderately
ionized  iron.  \citet{disalvo09}  identified in  the bright  atoll 4U
1705-44 resonant emission  lines of \ion{Ca}{xix}, \ion{Ar}{xviii} and
\ion{S}{xvi}, and an \ion{Fe}{xxv}  iron edge, which appears broad and
redshifted  with respect to  the expected  rest-frame energy.   It has
been therefore  suggested that  all these features,  and not  only the
broad iron  line, originate in  the reflection component.   To support
this interpretation, it has been shown that the smearing components of
the broad iron  line (i.e., the smearing parameters  of the reflection
component, the inner and outer radii, inclination angle of the system,
and  the  emissivity  index,  which  measures the  dependence  of  the
emissivity power-law  profile from the  distance to the source  of the
irradiating photons) consistently describe the shape and the broadness
of all the observed emission lines \citep{dai09,disalvo09,iaria09}.

In  this  work, we  perform  a  detailed  analysis of  two  XMM-Newton
observations  of the  bright atoll  source 4U  1705-44.  In  the first
observation, the  source was in a  low flux state, while  in the second
the  source  was  in a  bright  soft  state.  First results,  using  a
phenomenological  approach to  model the  reflection component  of the
latter observation  have been presented  in \citet{disalvo09}.  Here,
we  focus on  the spectral  changes  that occurred  between these  two
observations  using a self-consistent  reflection model.   

The  angular  dependence  of  the  reflected  component  formed  by  a
constant-density  partially-ionized medium  was originally  studied by
\citet{zycki94},  \citet{zycki94b}, and  \citet{matt03}.  Reprocessing
in  a   medium  in  hydrostatic   equilibrium  was  then   modeled  by
\citet{raymond93},      \citet{nayakshin00},     \citet{ballantyne01},
\citet{ballantyne02},     and     \citet{rozanska02}.     Furthermore,
\citet{nayakshin02}  examined the photoionized  accretion discs  via a
novel time-dependent  approach. A  hot layer forms  at the top  of the
disk atmosphere, roughly at  the inverse Compton temperature, followed
by a steep transition to colder, less ionized layers.

We     use    the    reflection     table    model     described    in
\citet{ballantyne04}.  First  application   of  this  table  model  to
consistently fit  the X-ray  spectrum of  an NS LMXB  can be  found in
\citet{ballantyne04b}.      This    reflection     model    (hereafter
\texttt{refbb})  is  calculated  for  an optically  thick  atmosphere,
irradiated   by   a  blackbody   incident   spectrum  of   $kT_{ion}$
temperature.  The model gives  the reflected spectrum according to the
ionization parameter  $\xi$, and the  relative abundance of  iron with
respect to the other metals. In addition to fully-ionized species, the
following     ions    are     included     in    the     calculations:
\ion{C}{iii}-\ion{C}{vi},                    \ion{N}{iii}-\ion{N}{vii},
\ion{O}{iii}-\ion{O}{viii},                  \ion{Ne}{iii}-\ion{Ne}{x},
\ion{Mg}{iii}-\ion{Mg}{xii},                \ion{Si}{iv}-\ion{Si}{xiv},
\ion{S}{iv}-\ion{S}{xvi},    and   \ion{Fe}{vi}-\ion{Fe}{xxvi}.    The
ionization parameter,  log $\xi$, can vary  between 1 and  4, with the
density n$_H$ of the illuminated slab constant at 10$^{18}$ cm$^{-3}$.
The  constant density prescription  can be  considered a  good diluted
approximation of the actual hydrostatic  structure in the disk for the
1.0-10.0 keV energy range \citep{ballantyne01}.

The $kT_{ion}$ temperature can vary between  1 keV and 5 keV.  All the
metals abundances  are fixed at the  solar value, except  for iron for
which models  were calculated for 0.1,  0.3, 1.0, 3, and  10 times the
solar value.  The space parameters  of this model covers, therefore, a
wide range of possible spectral solutions.  In particular, this is one
of the few  available reflection models in which  the primary incident
spectrum is  a soft thermal  spectrum, and, also  on the basis  of the
results already  shown in \citet{disalvo09},  it is the  most suitable
for fitting the soft states of NS LMXBs.

  We  show that,  within  the available  energy  range, the  continuum
  emission  can  be  simply  accounted  for  a  three-component  model
  composed   of  thermal   disk   emission,  a   saturated/unsaturated
  Comptonized harder  emission, and  a reflection component.  The last
  one  arises from  the disk  reflecting matter,  where  the impinging
  radiation  field \emph{is} the  hard X-ray  emission.  We  study the
  chemical  abundances of  the reflecting  disk matter,  the accretion
  flow in  the two states,  and possible scenarios for  explaining the
  spectral differences in the two states.

\subsection{\object{4U 1705-44}}

The source  is a  persistently bright, accreting  LMXB located  in the
direction of  the Galactic bulge  \citep{forman78}, hosting an  NS. It
shows  type-I X-ray  bursts, with  recurrence times  dependent  on the
accretion state \citep{langmeier87, gottwald89, galloway08}.  From the
peak  luminosity of  bursts  that exhibited  episodes of  photospheric
radius  expansions,  which are  thought  to  happen  at the  Eddington
luminosity   and   are,   therefore,   used   as   standard   candles,
\citet{haberl95} derived a  distance of 7.4$_{-1.1}^{+0.8}$ kpc, later
confirmed by  \citet{galloway08}.  The companion source  has still not
been identified,  although a near-infrared  counterpart, most probably
originated by  X-ray reprocessing by  the outer accretion  disk and/or
the companion star, has recently been found by \citet{homan09}.

\object{4U 1705-44} shows a secular trend toward alternating high- and low-flux
periods (see e.g., Fig.\ref{asm_countrate}) on a variable timescale of
months.   The  spectral variability,  on  the  contrary,  can be  much
faster, of days \citep{barret02}.  Classified by \citet{hasinger89} as
an Atoll  source, it  was later shown  that its spectral  and temporal
states are  intermediate between the  classic Atoll and  the Z-sources
division \citep{barret02}.   In particular, broadband  X-ray data have
shown that  the switch between hard  and soft states  can be explained
using a two-component model of a Comptonized inner emission and a soft
thermal emission. \citet{barret02} and \citet{olive03} interpreted the
alternate  hard and  soft  state transitions  as  caused by  different
truncation radii of  the accretion disk.  During hard  states the disk
is truncated  at a large  distance from the  compact object and  a hot
corona  with high electron  temperatures and  low optical  depth forms
around the NS.   During the hard to soft  state transitions, the inner
disk  approaches the NS;  this causes  an increase  in its  flux, thus
providing a more  efficient Compton cooling for the  hot electrons and
softer spectra.   This scenario is supported by  spectral modeling and
by  the  timing  analysis  of  the  power-density  spectra  where  the
characteristic  frequencies  of  the  band-limited noise  and  of  the
low-frequency  noise  components  are  strongly  correlated  with  the
position  of  the  source  in the  hardness-intensity  diagram.   High
frequency  quasi-periodic  oscillations   (QPOs)  are  also  observed,
usually in pairs (so-called kiloHerzt QPOs, kHzQPOs), with the highest
reported peak at 1160 Hz \citep{wijnands98, ford98}.

Spectral analysis with  the Chandra high-resolution gratings revealed,
superimposed  to the  continuum, a  set  of local  features, the  most
prominent of which was an unambigous, intrinsically broad (FWHM $\sim$
1.2 keV) fluorescent iron line \citep{disalvo05}.  However, it was not
possible to  distinguish among different broadening  mechanisms on the
basis of the goodness of the spectral fit. This motivated the need for
new observations with the  XMM-Newton satellite, given the much larger
collecting  area in  the iron  region of  the Epic-pn  CCDs.   A first
XMM-Newton observation caught  the source in a low  state, and the S/N
was rather poor in the iron range.  A second observation, performed as
a target of  opportunity when the source returned  to a high-intensity
soft period  was successful  in disclosing the  asymmetry in  the iron
line  shape,  which \citet{disalvo09}  interpreted  as  the result  of
reflection on  a disk surface, very  close to the NS,  of hard coronal
photons.  A similar scenario has also been proposed for BeppoSAX broad
band  data in  \citet{piraino07}, and  a  claim was  also made,  using
INTEGRAL  high  energy data,  of  a signature  of  a  Compton bump  in
\citet{fiocchi07}.  Recently,  \citet{reis09} using broad  band SUZAKU
data also shows  that the asymmetry of the  iron profile are naturally
described by a disk reflection scenario.

\section{Observation and data reduction}

The XMM-Newton Observatory \citep{jansen01} includes three 1500 cm$^2$
X-ray telescopes  each with an  European Photon Imaging  Camera (Epic,
0.1--15 keV) at  the focus. Two of the  Epic imaging spectrometers use
MOS  CCDs \citep{turner01},  and one  uses pn  CCDs \citep{struder01}.
Reflection     grating    spectrometers    (RGS,     0.35--2.5    keV,
\citealt{denherder01}) are located behind two of the telescopes.

   \begin{figure}[h!]
   \centering
   \resizebox{\hsize}{!}{\includegraphics[angle=-90]{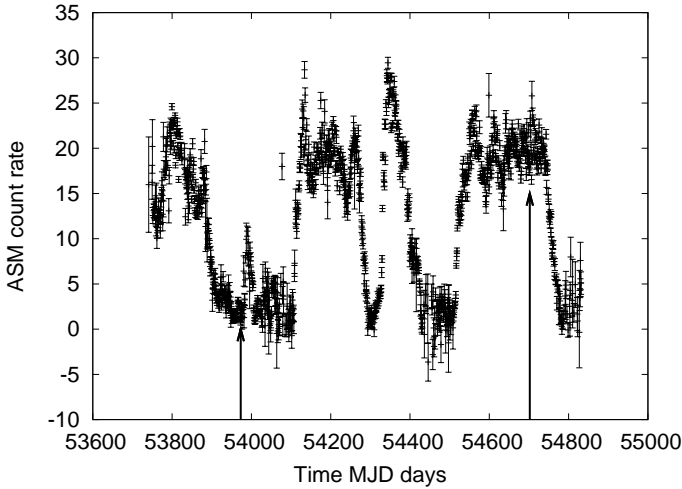}}
   \caption{\scriptsize{Long  term X-ray  light curve  of  \object{4U 1705-44}.
       Data  from the  All Sky  Monitor (2-10  keV range)  onboard the
       Rossi XTE,  from 1 January 2006  (MJD 53736) to  1 January 2009
       (MJD 54832). The two arrows in the figure mark the day when the
       SoftObs  (right  arrow)  and  the  HardObs  (left  arrow)  took
       place.}} \label{asm_countrate}%
    \end{figure}

XMM-Newton  observed 4U  1705-44 on  two occasions.   The  most recent
observation was performed on 24  August 2008 from 02:57:04 to 17:13:35
UTC (Obs. ID  0551270201, hereafter SoftObs) for a  duration of 51.390
ks and an effective exposure of 45.170 ks (owing to telemetry overflow
during  some part  of the  observation), the  second  observation took
place  on  26 August  2006  from 05:04:54  to  14:55:38  UTC (Obs.  ID
0402300201, hereafter HardObs), with  an exposure total time of 36.733
ks.

On both  occasions, the  Epic-pn operated in  timing mode  with medium
filter. In  this mode  only one central  CCD is  read out with  a time
resolution of  0.03 ms. This  provides a one-dimensional image  of the
source  with the  second spatial  dimension being  replaced  by timing
information. RGSs operated in spectroscopy mode.

\begin{figure}[htb!]
   \centering
   \includegraphics[angle=-90, width=8cm]{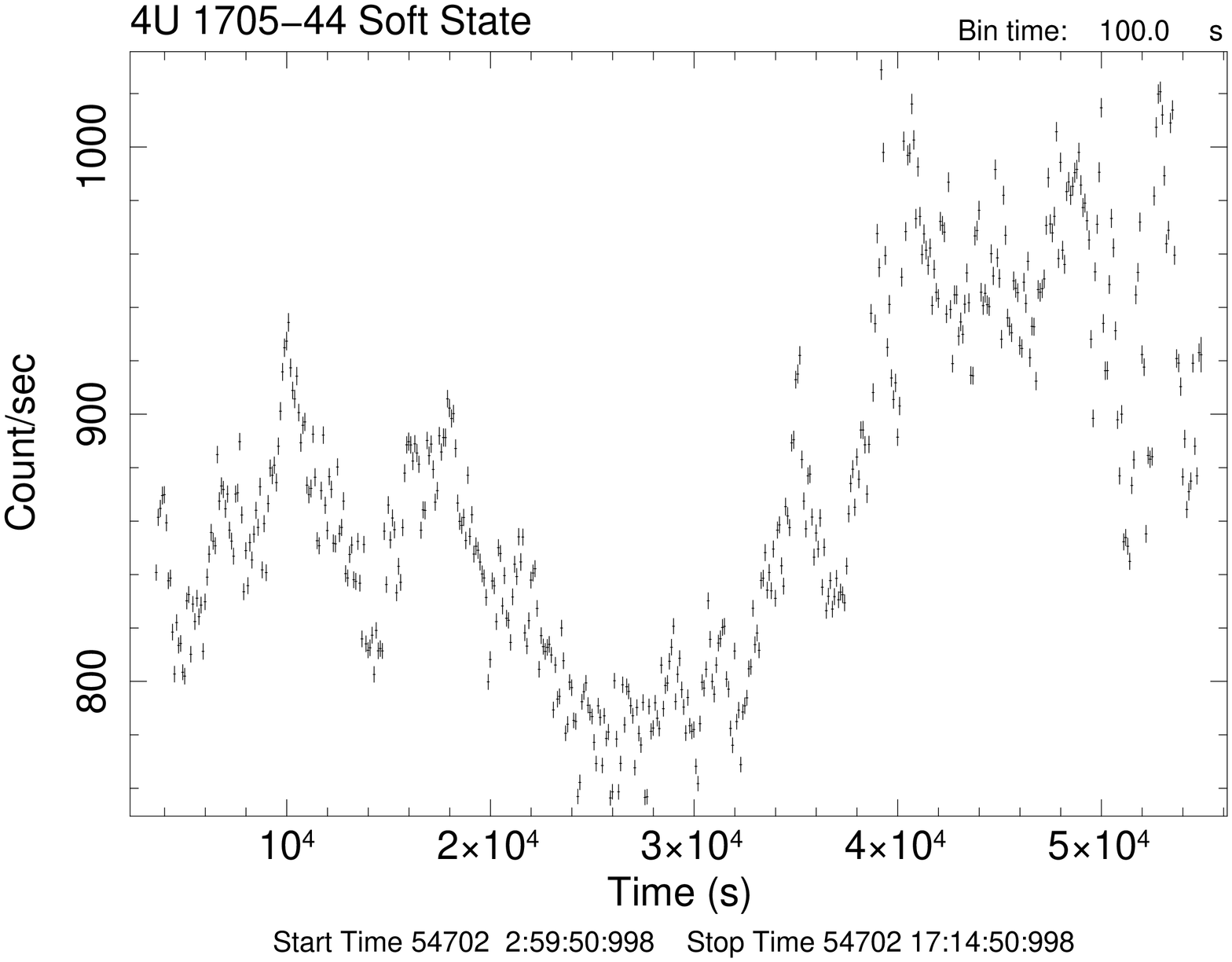}
   \includegraphics[angle=-90, width=8cm]{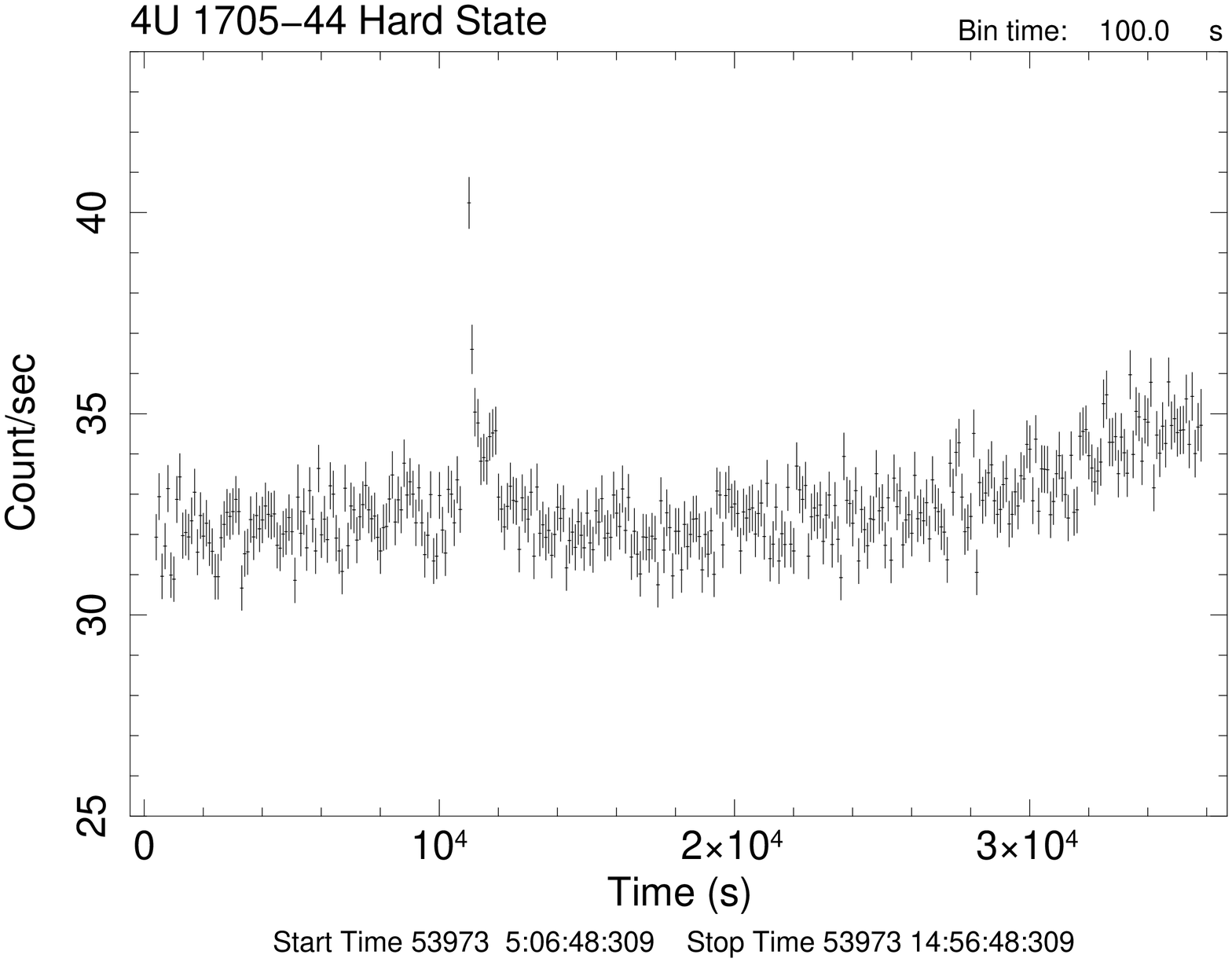}
   \caption{\scriptsize{Light curves for  SoftObs (upper panel) and for
       HardObs (lower panel).  In  HardObs, a type  I X-ray  burst is
       present at $\sim$ 11 ks  after the beginning of the observation
       (see Sect.\ref{burstanalysis} for details.)}}
              \label{light_curves}%
    \end{figure}

Figure  \ref{asm_countrate} shows the  long-term X-ray  variability of
the source  as seen from the All  Sky Monitor (ASM) of  the Rossi XTE,
from 1  January 2006 to  1 January 2009,  the period when  SoftObs and
HardObs were  performed is  marked in the  light curve.   HardObs took
place  when the source  activity was  in a  relatively low  state (ASM
count rate  $\sim$ 5 count/s),  while SoftObs was performed  almost at
the peak  of one of  the months lasting  active state (ASM  count rate
$\sim$ 35  count/s).  During  the HardObs the  source showed  a type-I
X-ray burst, while  no bursts were present in  the SoftObs.  In Fig. 
\ref{light_curves},  we  show the  Epic-pn  light  curves  of the  two
observations.  In  HardObs the  countrate variability,  excluding the
burst interval (see Sect.\ref{burstanalysis}),  is around a 10\% level
from the average of 33.4 counts/s, while SoftObs has an average of 870
counts/s and a similar variability spread.

\begin{figure}[htb!]
   \centering
   \resizebox{\hsize}{!}{\includegraphics[angle=-90]{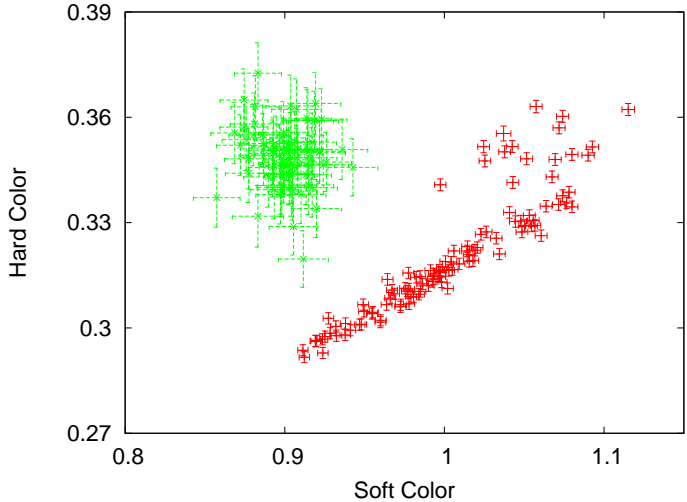}}
   \caption{\scriptsize{Color-color  diagram that shows  the relative
       positions  of the  two observations.   Each point  represents a
       time segment  of 500  s. SoftObs data  in red, HardObs  data in
       green.}}
              \label{colorcolor}%
    \end{figure}

    In Fig. \ref{colorcolor}, we  show the color-color  diagram for
    the two observations.   The soft color is defined  as the ratio of
    the counts in the 1-3/3-6 keV band and the hard color as the ratio
    in  the   3-6/6-10  keV  band.   The  source   shows  very  little
    variability during  HardObs, while  SoftObs is more  variable, but
    always softer than HardObs.

    For   spectral    analysis   we   used    Xspec   version   12.5.0
    \citep{arnaud96}.   The  interstellar  absorption  (\texttt{phabs}
    component in Xspec) is calculated using the cross-section table of
    \citet{verner96}    and   the    metal    abundances   table    of
    \citet{wilms00}.   We  used the  Science  Analysis System  ($SAS$)
    version  9.0 for  the reprocessing  of XMM-Newton  data,  with the
    $epproc$ pipeline  for generating the  Epic-pn scientific products
    and the $rgsproc$ pipeline for the RGS ones.

    We used standard $SAS$ tools for data selection, masks generation,
    ancillary and response  matrices.  Spectra were filtered according
    to the $PATTERN \leq 4$ and $FLAG=0$ criteria.  Background spectra
    above 12  keV were  inspected to check  for periods of  high solar
    background.   Background  spectra were  extracted  using the  same
    filtering  criteria of source  spectra in  the RAWX  3-10 columns.
    Source spectra were extracted in  the RAWX 28--46 columns. We used
    the  Epic-pn  spectra  grouped  to  have  5  energy  channels  per
    resolution  element and  at least  25 counts  per  energy channel.
    This choice avoids an  excessive oversampling of the data, without
    loss  of  important  spectral  information.  RGS  data  were  also
    grouped by adopting the 25 counts per energy channel criterion.

    By  testing different choices  of channels  binning, we  observed a
    change in  the value of  the reduced $\chi^2$ of  the best-fitting
    models,   but   still   the   best-fitting  parameters   are   not
    significantly   dependent  on   this   choice.

\section{Spectral Analysis}

\subsection{SoftObs \emph{The soft state}}

SoftObs  was   analyzed  in  \citet{disalvo09};  here   we  present  a
self-consistent approach  to the  spectrum modeling using  a broadband
reflection  model.  In \citet{disalvo09},  the main  focus was  on the
determination and the  nature of the broad emission  lines seen in the
energy  spectrum from  2.5  keV to  11  keV.  It  was  shown that  the
reflection scenario,  including the relativistic effects,  is the most
physically  and statistically  favored solution  for  interpreting the
broad emission features.

We analyzed  these data again,  using a slightly  different extraction
mask  for  the Epic-pn,  namely  excluding  the  brightest CCD  column
(RAWX=37), in  order to obtain  a complete pile-up free  spectrum.  We
used  the  $epatplot$   tool  of  the  $SAS$  package   to  check  the
distribution of  single vs.  double event  pattern distribution.  This
tool shows the theoretical normalized fractional distribution of these
events and  the real data  distribution.  A mismatch between  the data
and the theoretical predicted  curves indicates possible pile-up.  The
XMM-Newton  data  of 4U  1705-44,  including  the  brightest CCD  row,
deviate from  the theoretical  curves for energies  less than  2.5 keV
(which   indeed  were   excluded  from   the  analysis   published  by
\citealt{disalvo09}) and  above 9 keV. The exclusion  of the brightest
row of  the Epic-pn CCD  significantly reduces this problem  (e.g. the
gap between  data and  theoretical curve passes  from 0.02 to  0, from
0.03  to $<$0.01,  from  0.04  to $<$0.02  at  3 keV,  6  keV, 9  keV,
respectively).

\begin{figure}[htb!]
   \centering
    \resizebox{\hsize}{!}{\includegraphics[angle=-90]{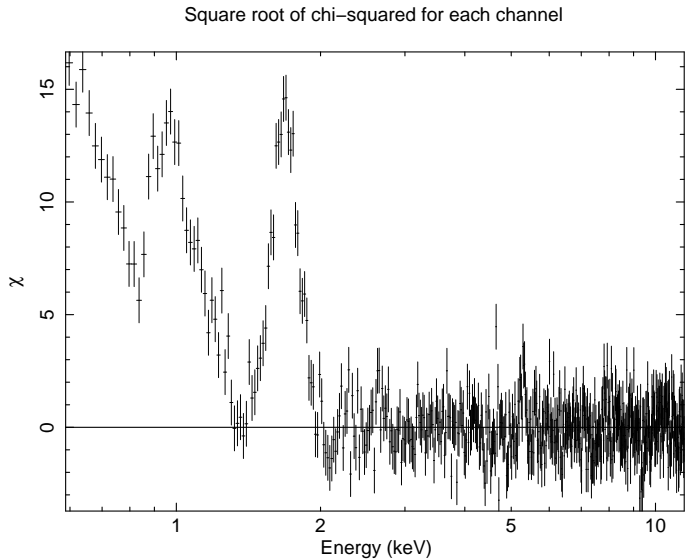}}
   \caption{\scriptsize{Residuals, in units  of $\sigma$, in the broad
       0.6-12  keV energy  range,  of the  Epic-pn  spectrum when  the
       2.0-12.0    keV    best-fitting    model   is    used    (Table
       \ref{fits_1705soft}).  A  broad feature  at 1.7 keV  is clearly
       visible  together  with soft  residual  emission  below 1.0
       keV.}}
              \label{residuals_softpn}
    \end{figure}

We also  used RGS1 data  in the 1.0-2.0  keV energy range,  while RGS2
data   have   been  discarded   since   they   are  severely   pile-up
affected\footnote{The RGS2  frame time is  double with respect  to the
  RGS1 CCDs caused by an instrumental failure occurred in August 2007;
  the longer  frame time  makes the RGS2  CCDs more easily  subject to
  pile-up.}. Epic-pn  data were  first used in  the broad  energy range
between 0.6-12.0  keV.  This range, however, revealed  (as in HardObs)
some very large systematic features  around the Au and Si instrumental
edges  that are  difficult to  reconcile with  any  plausible physical
scenario  \citep[see also][]{boirin03,  papitto09}.  Also,  a mismatch
between the RGS and PN spectrum in the 1.0-2.0 keV range is indicative
of calibration uncertainties that affect the determination of the soft
band.  In Fig.\ref{residuals_softpn}, we show the Epic-pn spectrum for
clarity,  when  the   2.0-11  keV  band  is  used   for  deriving  the
best-fitting model, and then the 0.6-2.0 keV channels are noticed. The
broad feature at 1.7 keV is also present in HardObs, and we found that
this spectral shape  can be well-fitted using a  simple broad Gaussian
line. These Gaussian parameters are consistent in the two datasets. At
very  low energies,  the spectrum  shows a  clear flux  excess, which,
however, is ruled out by  the simultaneous RGS1 dataset.  We therefore
excluded the PN  energy range below 2  keV in order to get  rid of any
systematic problem that  could affect our analysis, and  used only the
RGS1 data to constrain the softest band.

    In   analyzing  these   data,  \citet{disalvo09}   decomposed  the
    continuum emission into the sum  of a soft blackbody component and
    a harder, optically thick,  Comptonized component.  In the present
    work,   we  used   a   slightly  different   but  physically   and
    statistically  similar  continuum  model.   We replaced  the  soft
    blackbody   component    by   a   multicolored    disk   component
    (\texttt{diskbb} model in Xspec, \citealt{makishima86}), while the
    Comptonized  emission could  be approximated  well with  a simpler
    hard   blackbody  component.    Having  an   additional  continuum
    component at  energies above 7  keV from the  reflection continuum
    gives, in fact, a Comptonization  optical depth over 10, while, at
    the  same time, seed-photon  temperature and  electron temperature
    get very near to  each other ($\Delta E < 0.5 $  keV), so that the
    blackbody      approximation      is     reasonably      justified
    \citep{dai09,iaria09}.   Moreover, instead  of a  number  of local
    \texttt{disklines} and a smeared  absorption edge, in this work we
    use a self-consistent table reflection model \citep{ballantyne04},
    convolved with  the \texttt{rdblur} component  \citep{fabian89} to
    consider  the relativistic effects  in a  Schwarzschild space-time
    (i.e.  relativistic  Doppler boosting, light  bending, and lensing
    effect of the  disk irradiated matter). The degree  of smearing is
    set  by   the  disk's  inner   and  outer  radius  (in   units  of
    gravitational radii R$_g$  = $GM/c^{2} \sim 2$ km  for a 1.4 solar
    mass NS),  the inclination angle  of the disk, and  the emissivity
    law profile, assumed to only  depend on the distance of the source
    of  the  impinging  photons  ($\epsilon  (r)  =  r^{-q}$).   These
    parameters  are well constrained  by the  fitting, except  for the
    disk  outer  radius,  which  is poorly  constrained  (R$_{out}$  =
    10$_{-6}^{+100}$ $\times$ 10$^3$ R$_g$), so that we choose to keep
    it  frozen at 10$^{4}$  R$_g$, given  its much  larger uncertainty
    with respect to the other spectral parameters.

    We tested that the  $kT_{ion}$, the temperature of the irradiating
    flux is perfectly consistent with the same temperature of the hard
    blackbody component,  and we set  the two spectral  parameters tied
    together during the fitting procedure.

\begin{figure*}[h!]
   \centering
   \includegraphics[angle=-90,width=8cm]{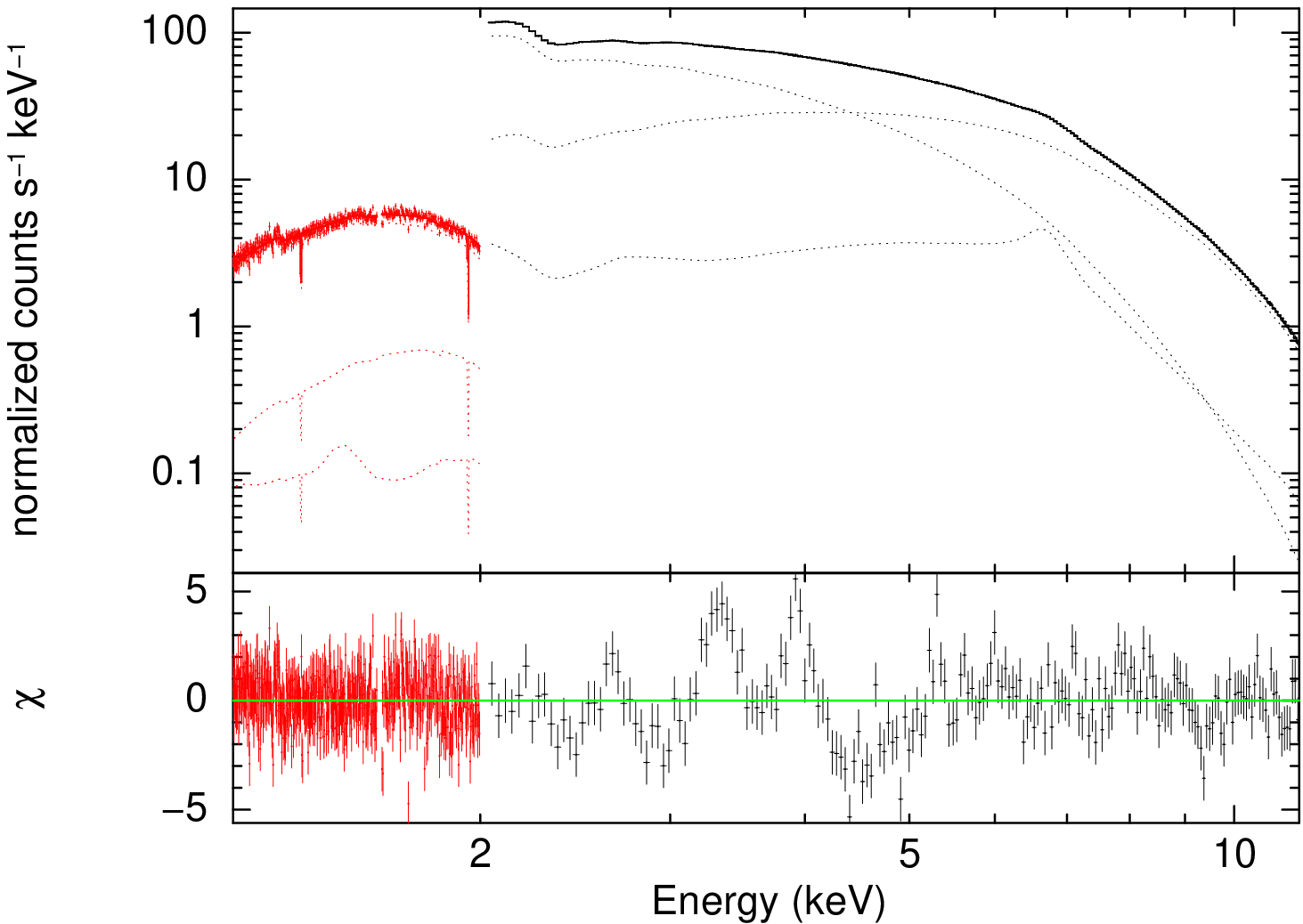}
   \includegraphics[angle=-90,width=8cm]{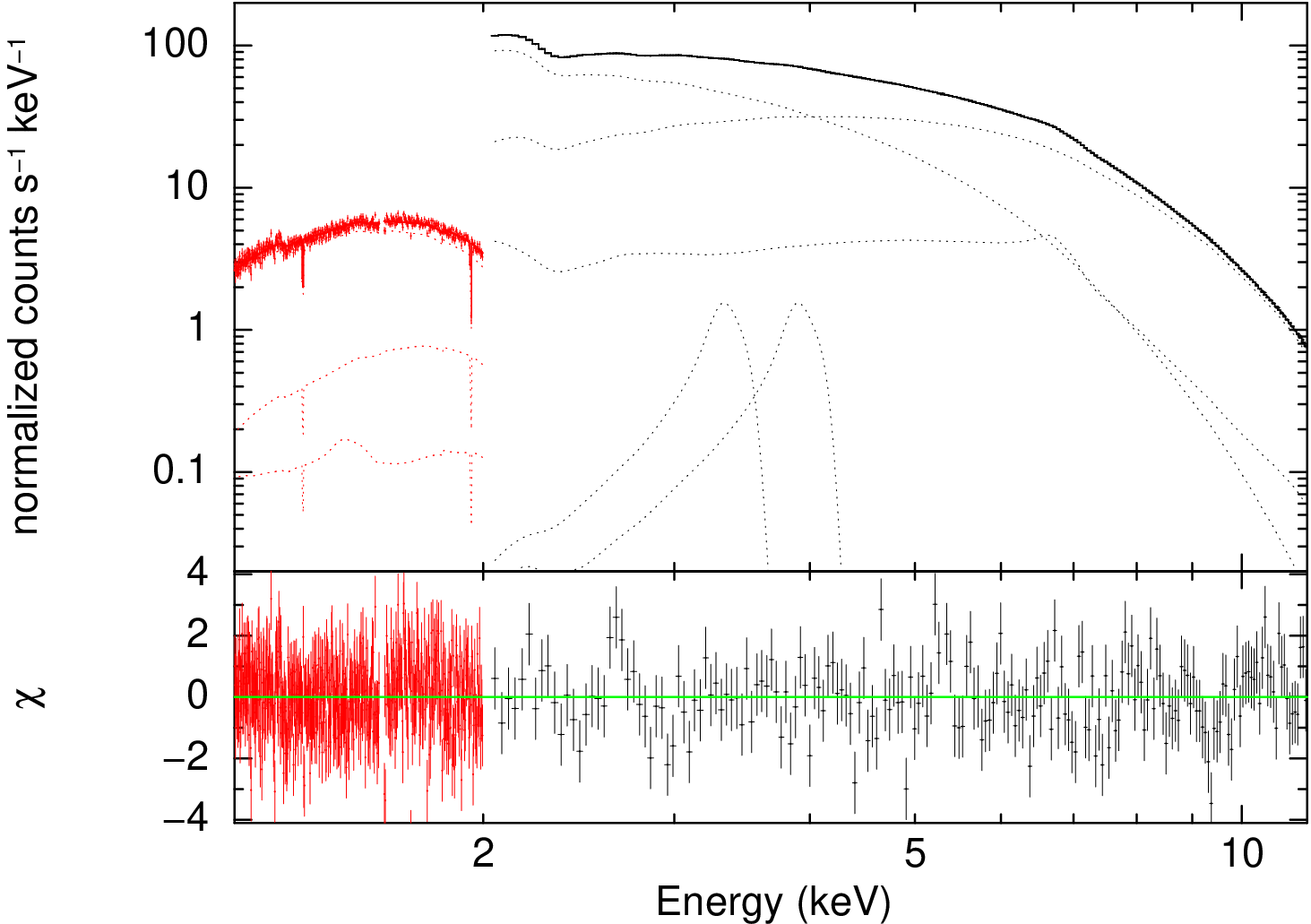}
   \caption{\scriptsize{Data  and best-fitting  model  with individual
       components  separately   shown  (top  panels),   together  with
       residuals in  units of sigma  (lower panels).  Left  panel: the
       best-fitting    model   without    the    \ion{Ar}{xviii}   and
       \ion{Ca}{xix} emission  lines. Right panel:  data and residuals
       using the best-fitting model of Table \ref{fits_1705soft}.}}
              \label{residuals_pn05}%
    \end{figure*}

    This model adequately describes the continuum and the shape of the
    iron line,  and also takes the emission  features of \ion{Si}{xiv}
    and \ion{S}{xvi}  into account, although some  residuals are still
    present at $\sim$ 2.6 keV probably due to an overabundance of this
    metal with  respect to  the other elements.   There are  still two
    broad features  at $\sim$ 3.9 keV  and $\sim$ 3.32  keV that stand
    out in  the residuals (see  Fig.\ref{residuals_pn05}, left panel).
    Following the interpretation  of \citet{disalvo09}, which ascribes
    these features  to resonant emission lines of  He-like calcium and
    H-like argon  ions, we  added two Gaussians  to account  for these
    residuals  profiles, given  that  our reflection  model lacks,  in
    fact, the relative emission  lines for these elements. These lines
    are  kept with  0  keV frozen  width  and are  convolved with  the
    \texttt{rdblur} kernel, with  the smearing parameters identical to
    the  \texttt{refbb}   model.   The  reduced   $\chi^2$,  with  the
    introduction  of these  two additional  smeared lines  varies from
    1.70 (710 dof) to 1.30 (706 dof), and the residuals at the 3.3 keV
    and    3.9    keV   energies    are    reasonably   well    fitted
    (Fig. \ref{residuals_pn05},  right panel). These two  lines have a
    similar  equivalent width  (between 8  and  10 eV),  and the  line
    positions are well constrained  and in agreement with the expected
    rest-frame  emission line  energies (the  \ion{Ar}{xviii}  at 3.30
    $\pm$ 0.02 keV and the \ion{Ca}{xix} at 3.90 $\pm$ 0.02 keV).

    Given the high quality of  the spectrum, we also tried to evaluate
    the fraction of  iron abundance with respect to  the other metals.
    Setting  the  iron fraction  overabundant  with  respect to  other
    metals  by a  factor of  3 provides  a worsening  of the  fit with
    respect  to the  assumed solar  values  ($\Delta \chi^2$  = 10,  F
    statistic value  9.7 and probability 0.2\%), and  the same happens
    when the iron fraction is assumed to be underabundant for the same
    factor  ($\Delta  \chi^2$  =   12,  F  statistic  value  11.7  and
    probability 0.06\%).  We concluded  that the iron fraction is well
    constrained  by  the  reflection   model,  and  it  is  relatively
    consistent with the assumed solar abundance values within a factor
    of 3.

    Although  this model is  formally not  satisfactory ($\chi^2_{red}
    \sim 1.3$ with 706 dof), there  is no evidence in the residuals of
    other  local   features,  and   the  residuals  are   so  randomly
    distributed around the best-fitting  model that no other continuum
    component  is required.   The most  probable explanation  for this
    high $\chi^2$ is  that, at this very high count  rate ($>$ 400 cts
    s$^{-1}$),  the statistical  errors  in the  channels becomes  too
    small  compared  with the  relative  uncertainty  of the  response
    matrix, giving an overall  underestimation of the systematic error
    to  be  attributed  to   the  energy  channels.   Adding  a  0.8\%
    systematic error  to the  best-fitting model, we  find a  value of
    $\chi^2$ $\sim$ 1.0.   This error is added in  quadrature in order
    to self-consistently  evaluate the error  to be associated  to the
    spectral    parameters,    which    are    reported    in    Table
    \ref{fits_1705soft}. We also report  the unabsorbed fluxes of each
    spectral component in the 1-10  keV range, and in the extrapolated
    0.1-100.0 keV  range.  The blackbody radius and  inner disk radius
    are calculated assuming a distance to  the source of 7.4 kpc and a
    disk inclination of 35$^{\circ}$.

\begin{figure*}[ht!]
    \centering
   \includegraphics[angle=-90,width=8cm]{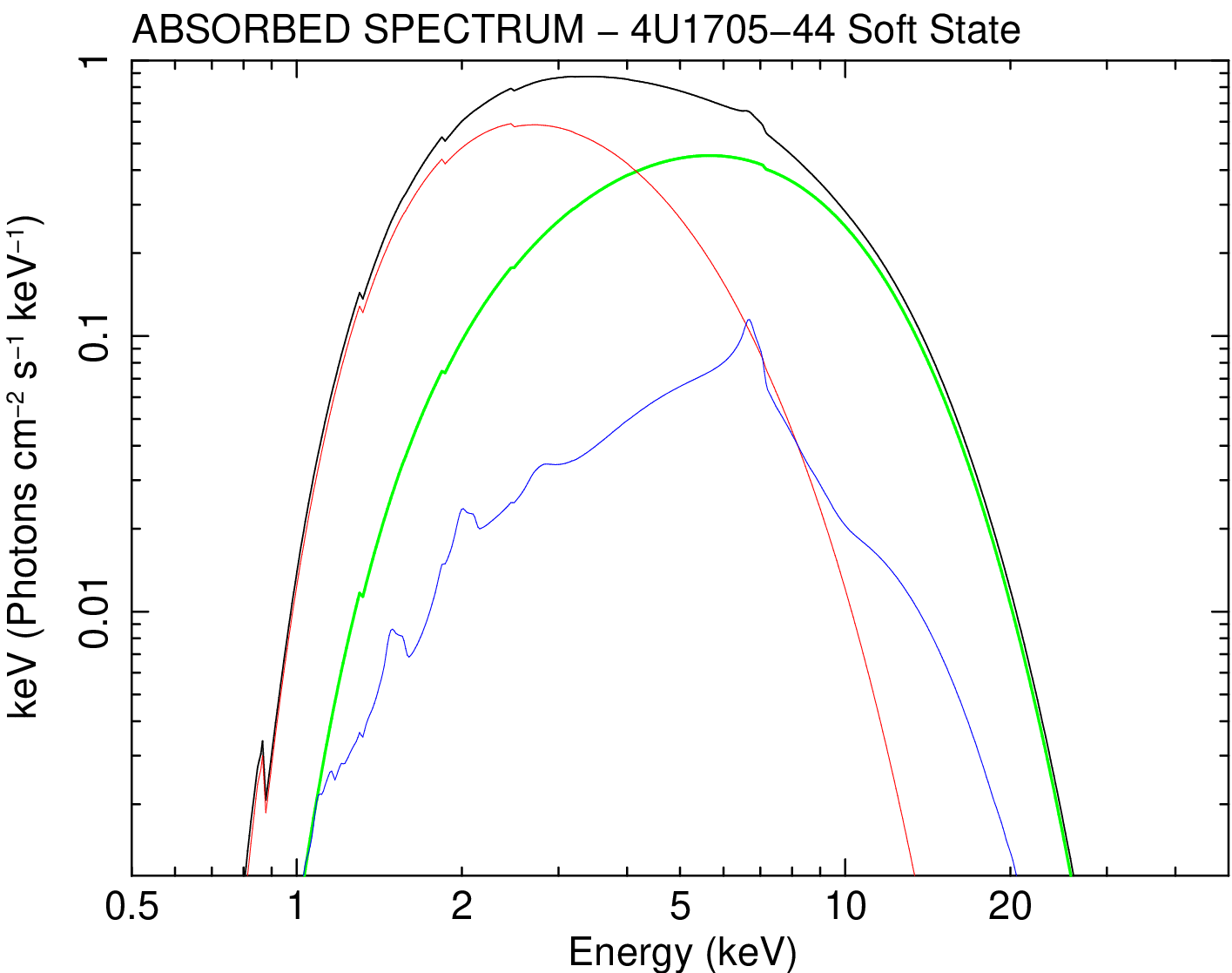}
   \includegraphics[angle=-90,width=8cm]{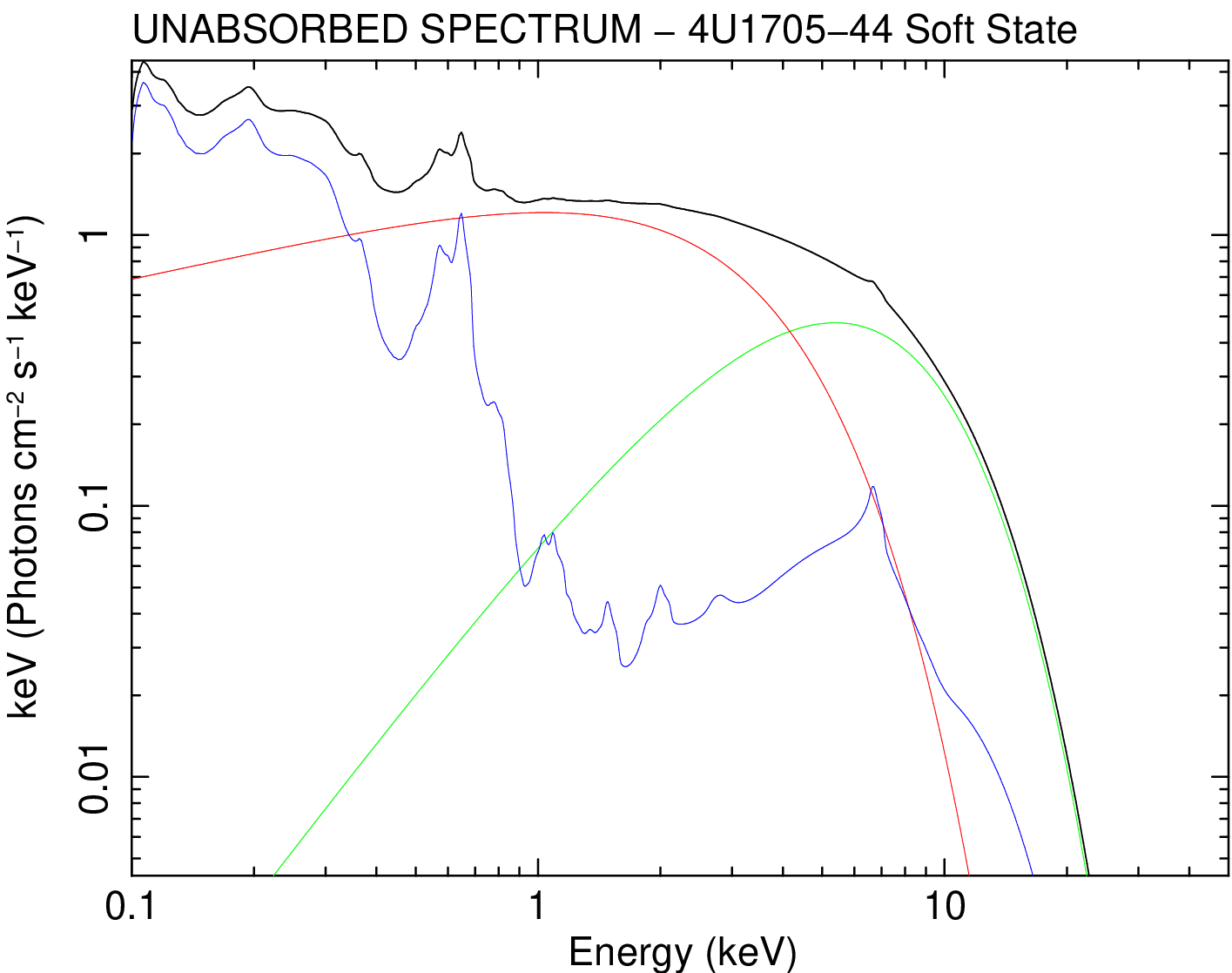}
   \caption{\scriptsize{Best  fitting  model in  the  E$\times$f(E)
     representation of the soft state of 4U  1705-44. Red line  is the
     thermal  disk  emission,  green   line  the  black  body  thermal
     emission, blue line the  reflection component. Left panel shows the ISM absorbed model, right  panel shows
     the unabsorbed model.}}
              \label{1705soft_emo}%
    \end{figure*}

    The observed 1.0-10.0 keV absorbed flux in this observation is 5.6
    $\times$  10$^{-9}$ erg cm$^{-2}$  s$^{-1}$, while  the unabsorbed
    flux     is    1.2     $\times$     10$^{-8}$    erg     cm$^{-2}$
    s$^{-1}$.  Extrapolation of  this model  from 0.1  keV to  100 keV
    results in a bolometric  unabsorbed flux of 1.6 $\times$ 10$^{-8}$
    erg  cm$^{-2}$ s$^{-1}$.   The inferred  X-ray luminosity  in this
    state  corresponds to  $\sim$ 1  $\times$ 10$^{38}$  erg s$^{-1}$.
    Figure \ref{1705soft_emo}  shows the contribution  of disk thermal
    emission, hard  thermal emission,  and reflection component  to the
    absorbed 0.5-50 keV flux (left panel) and to the unabsorbed 0.1-50
    keV  flux  (right  panel).   The contribution  of  the  reflection
    component is  dominant at very  low energies, below 1  keV, while,
    because of the very soft nature of the incident spectrum, it falls
    off very rapidly beyond the iron line region.


\begin{table}[htb!]
\caption{Best-fitting values  of the parameters obtained  for the soft
  state of \object{4U 1705-44}.}
\label{fits_1705soft}
\centering      
\begin{tabular}{ll c }
\hline\hline             
Parameter      & Units                           & Values \\
\hline                               
$N_\mathrm{H}$          & 10$^{22}$ cm$^{-2}$      &  2.70$\pm$0.02             \\     

Disk kT        & keV                              &  1.15$\pm$0.03         \\
Disk radius    & km                               &  11.3$\pm$0.5  \\
Disk flux      & 10$^{-9}$ erg cm$^{-2}$ s$^{-1}$  &  5.59-7.17  \\

BB kT          & keV                              &  1.906$_{-0.06}^{+0.20}$   \\
BB Radius      & km                               &  5.3$\pm$0.2 \\
BB flux        & 10$^{-9}$ erg cm$^{-2}$ s$^{-1}$  &  5.14-6.70 \\

Inner radius   & R$_g$                            & 6$^{+7}$  \\
Outer radius   & R$_g$                            & 10$^4$  \\
Betor          &                                  & 2.30$^{+0.08}_{-0.11}$  \\
Inclination    & degrees                          & 35.0$\pm$2.0  \\

log $\xi$      &                                  & 2.36$_{-0.18}^{+0.33}$  \\
Refbb Flux     & 10$^{-9}$ erg cm$^{-2}$ s$^{-1}$  & 0.81-2.21  \\

$\chi^2_{red}$ (d.o.f.)                            & & 726 (706)             \\
\hline
\end{tabular}
\end{table}


\newpage

\subsection{HardObs \emph{The hard state}}

In this  section we describe  the spectral analysis of  the persistent
emission  of HardObs,  excluding the  data close  to the  type-I X-ray
burst (from 15 s before the burst peak to 200 s after the burst peak).
We first used  Epic-pn data in the broad energy range  from 0.6 keV to
12.0 keV,  RGS1, and  RGS2 data  in the 1.0-2.0  keV energy  range.  A
single   absorbed  power-law   model  fails   to  describe   the  data
($\chi^2_{red}$ $\sim$  4.07, 826 dof);  a power law  with exponential
cut-off   provides   a    better   fit,   but   still   unsatisfactory
($\chi^2_{red}$  $\sim$ 2.82,  825 dof);  replacing  the exponentially
cut-off   power    law   with   a    physical   Comptonization   model
(\texttt{comptt} in  Xspec, \citealt{titarchuk94}) again  improves the
fit significantly ($\chi^2_{red}$ $\sim$ 1.61, 824 dof).  Residuals at
low  energies, however,  require the  addition of  an  additional soft
component.   A  blackbody  component  or  multicolored  disk  emission
(\texttt{diskbb}  in  Xspec)  is  able  to  satisfactorily  model  the
continuum emission, although from a  statistical point of view, we are
not able to  distinguish between a single thermal  temperature and the
multicolored  model  ($\chi^2_{red}$ $\sim$  1.40,  822  dof for  both
models), and  we discuss different physical scenarios.  The still high
$\chi^2$ stems from the presence of a broad local emission line in the
iron  range and  to an  S-shaped pattern  in the  residuals  at softer
energies, which could be fitted by a broad Gaussian at energies $\sim$
1.7 keV.  This broad line  is, however, not easily identified with any
physically reasonable bound-bound transition  so we discuss a possible
instrumental origin.   In fact, the residuals pattern  between 1.0 and
2.0  keV is  mostly guided  by the  Epic-pn channels,  given  the much
larger collecting area compared to the RGSs.  Fitting the RGS datasets
alone, we obtain a relatively good fit (reduced $\chi^2$ = 0.99) using
either an  absorbed power-law model  or blackbody model  as continuum.
RGS data alone indicate a  higher $N_\mathrm{H}$ than the Epic-pn data
($N_\mathrm{H}$ 3.0$\pm$0.4 $\times $10$^{22}$ cm$^{-2}$ for power-law
continuum,  2.2$\pm$0.3  $\times$  10$^{22}$ cm$^{-2}$  for  blackbody
emission).

\begin{figure*}[ht!]
   \centering
   \includegraphics[angle=-90,width=8cm]{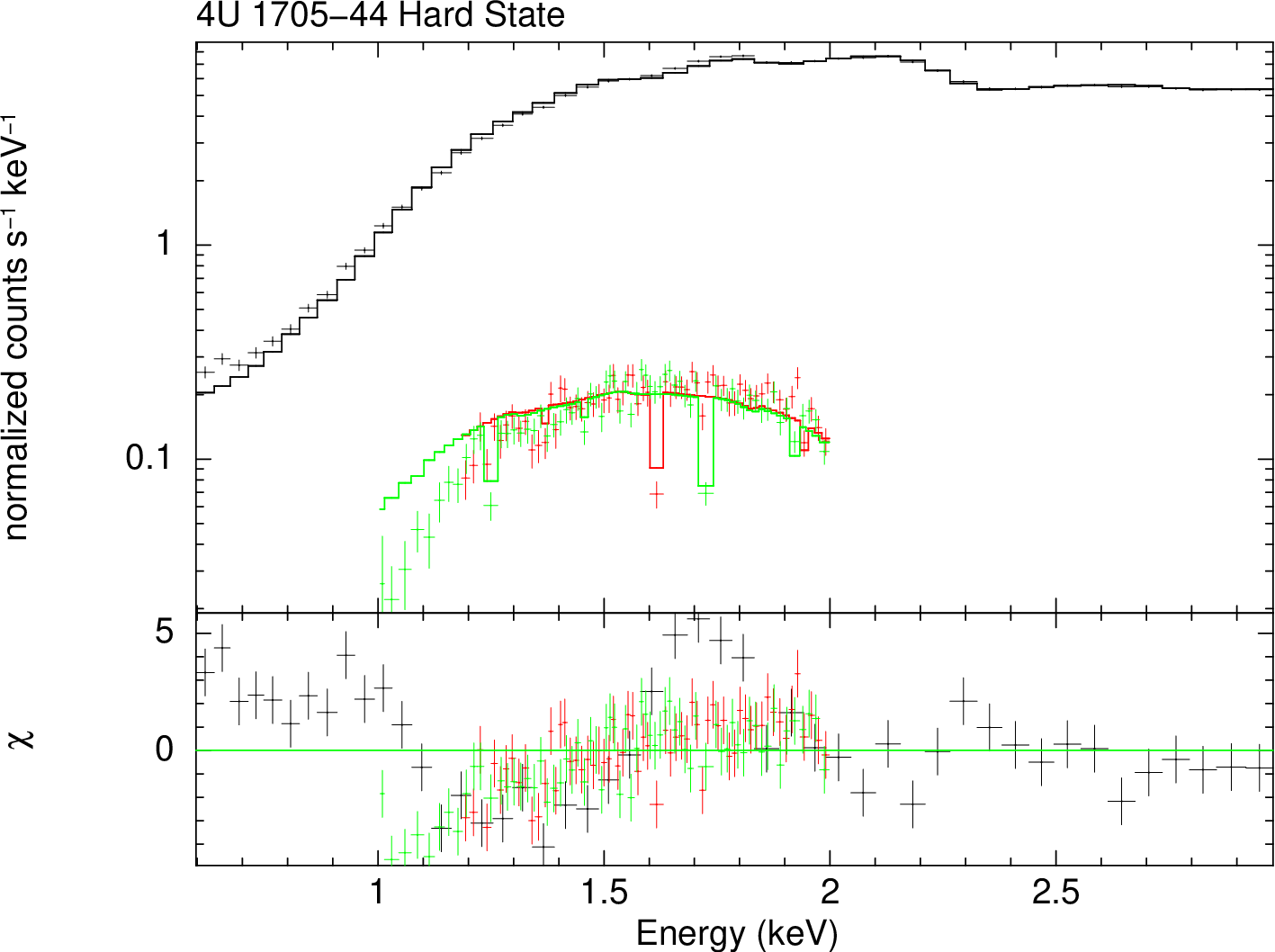}
   \includegraphics[angle=-90,width=8cm]{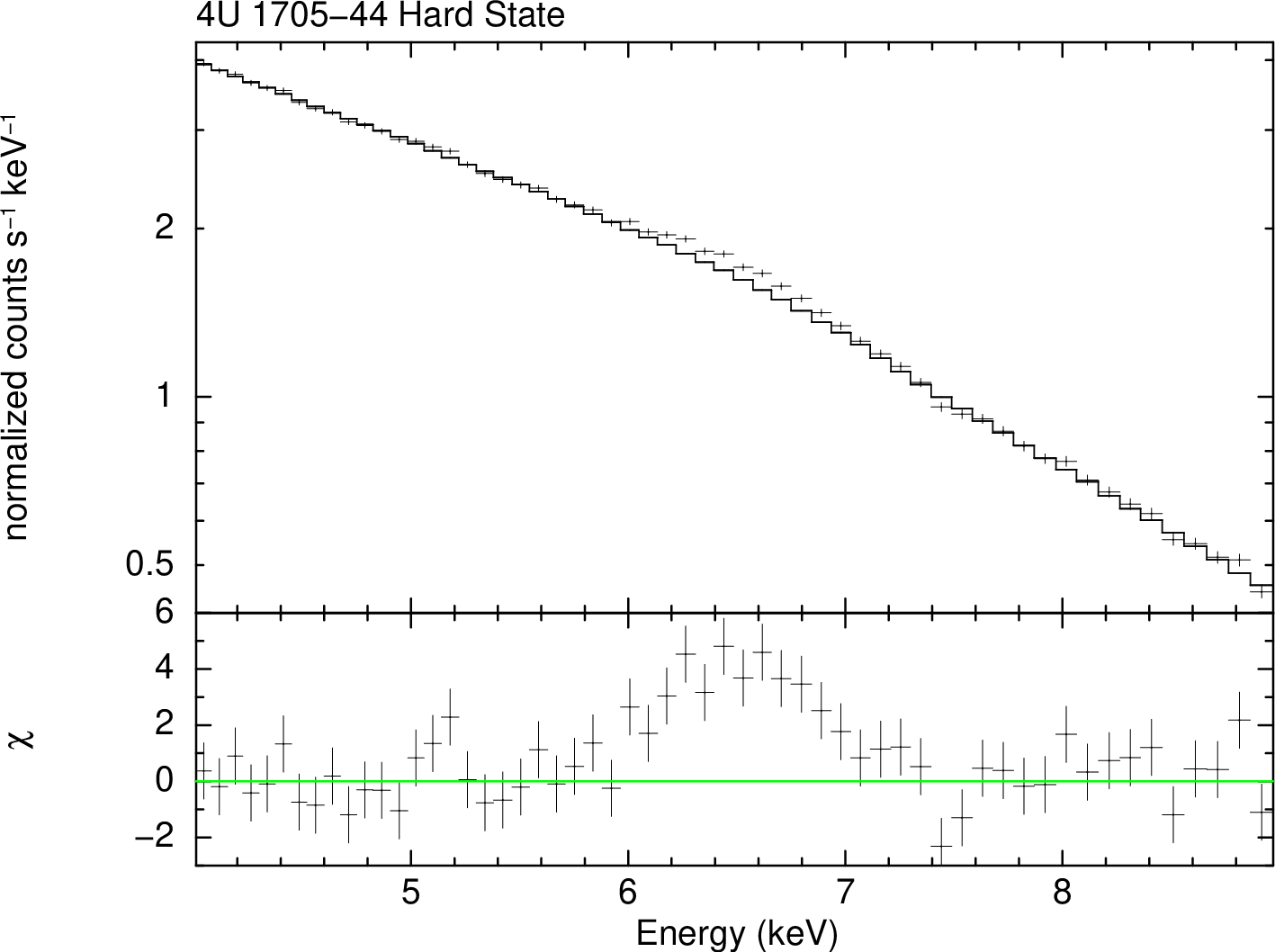}
   \caption{\scriptsize{Left  panel: data  and residuals  at  the soft
       energies   (0.6-3.0   keV),   using  an   absorbed   blackbody
       component. Residuals clearly display an S-shaped pattern in the
       Epic-pn data  (black data)  between 1.5 keV  and 2.0 keV  and a
       general  mismatch between  RGS data  (green and  red  data) and
       Epic-pn data  at softer energies.  Data have been  rebinned for
       clarity.  Right  panel: residuals in the iron  range, when data
       4.0-9.0  keV are  fitted  with Model  1  and the  iron line  is
       subtracted from the data.}}
              \label{1705hard}
    \end{figure*}

On  the contrary,  in  the  restricted 0.6-2.0  keV  Epic-pn band,  we
clearly observe large  residuals that cannot be ascribed  to the shape
of  the continuum  emission  (see Fig.   \ref{1705hard}, left  panel).
Inclusion  of a  broad line  at 1.7  keV makes  the  model acceptable;
however, this line is not required  in the RGSs spectrum, for which we
get  only an  upper  limit on  the  line normalization  of 8  $\times$
10$^{-5}$  photons cm$^{-2}$ s$^{-1}$  (90\% confidence  level), which
not  compatible with  the lower  limit  of the  Epic-pn spectrum  (2.9
$\times$  10$^{-4}$  photons  cm$^{-2}$  s$^{-1}$  ,  90\%  confidence
level).  This led us to conclude that the line is probably an artifact
due to incorrect modelization of the  instrumental K Si and M Au edges
of the Epic-pn. Moreover, as mentioned above, the Epic-pn data require
a  much  lower  value  of  the  $N_\mathrm{H}$  (1.3$\pm$0.2  $\times$
10$^{22}$ cm$^{-2}$),  which does not match  by far the  high value as
derived  from  the RGSs  and  the  Epic-pn data  above  2  keV or  the
$N_\mathrm{H}$  value  of SoftObs,  nor  the  $N_\mathrm{H}$ value  as
derived  from  the  burst  analysis.   This implies  either  an  usual
underabundance of light Z  metals (e.g., oxygen, neon, magnesium) with
respect  to high Z  metals or  the presence  of additional  soft X-ray
emission, or a soft flux miscalibration of the instruments.  The first
hypothesis appears  the least probable. The second  hypothesis can be,
in principle, correct and it  is not difficult to image other physical
mechanisms that  can play an important  role at these  very soft X-ray
energies (X-ray  reprocessing in the outer  disk or at  the surface of
the  companion  star),  but   our  experience  has  shown  that  other
XMM-Newton observations  in timing mode of  moderately absorbed bright
sources  are affected  by such  soft excesses  \citep{dai09, iaria09},
often in the form of strong  and very broad emission lines at 1.0 keV.
Although  possibly  hiding  a   physical  origin,  such  excesses  can
introduce strong  biases in the determination of  the correct physical
model  and  physical  conclusions.   In view  of  the  above-mentioned
considerations, we chose not to  consider Epic-pn data below 2 keV, in
analogy with  data reduction of SoftObs,  and to rely on  the RGS data
for the rest of our analysis.

RGS data, however, have fewer statistics than the Epic-pn data, and we
noted  that  the  determination  of  the soft  component  is  strongly
correlated with the $N_\mathrm{H}$ value.   We chose to use results of
the  soft  state  analysis  to  derive  better  constraints  for  this
observation, and we froze the $N_\mathrm{H}$ value to the best-fitting
value   of   Table  \ref{fits_1705soft}.    Adopting   the  value   of
2.44$\times$10$^{22}$  $\times$  cm$^{-2}$   derived  from  the  burst
analysis (see Sect.\ref{burstanalysis})  does not significantly change
the determination of the spectral components, or our conclusions.

    The broad $\sim$ 6.4 keV  line can be identified as a fluorescence
    emission line  of iron in a  low ionization state (see  Model 1 in
    Table  \ref{fits_1705hard}).   The  line is  intrinsically  broad,
    given that its width is  much greater than the energy resolution of
    the Epic-pn in this energy band.

    We  present  in Table  \ref{fits_1705hard}  the  results of  three
    different spectral decompositions. The models mostly differ in the
    choice  of the  soft component  and of  the reflection  model, but
    share the use of a  Comptonized component for the hard-energy part
    of the spectrum.  However, because the spectrum is only covered up
    to 12  keV and  the Comptonized emission  extends well  above this
    energy,  we found  a strong  correlation,  in all  the models,  in
    parameter space, between the  electron temperature and the optical
    depth  of this  component.  In  this  state, the  source has  been
    previously observed many times with the Rossi XTE satellite, whose
    higher  energy coverage  allows for  much better  constraints.  In
    \citet{barret02}, the optical  depth of the Comptonized component,
    in  this hard  state, shows  a nearly  constant value  at  5.5. We
    chose, therefore, to keep this  parameter frozen at this value, to
    reduce  the uncertainties in  the determination  of all  the other
    spectral  parameters, after  having  checked that  this value  is,
    however, within  the parameter error bar,  if this is  set free to
    vary in the fitting procedure.

      In Model 1,  we use a thermal blackbody  for the soft component
      and  a Comptonization  for the  hard  component \citep{barret02,
        gierlinski02}.   The  iron line  is  fitted  using a  Gaussian
      line. In Model 2, we use a multicolored disk emission, with disk
      normalization  $N$ fixed  to 70  (this value  corresponds  to an
      inner   edge   located  at   $\sim$   1.2   R$_{NS}$)  and   the
      \texttt{refbb}  reflection model convolved  with \texttt{rdblur}
      smearing  kernel.  The use  of the  \texttt{refbb} model  can be
      considered a good first-order approximation, within the 1-12 keV
      available band, of the  actual reflection spectrum, although the
      impinging  radiation  spectrum  now  departs  from  a  saturated
      blackbody-like spectrum. The reflection  model is mainly used to
      reproduce the broad  iron fluorescence line as we  have no other
      prominent  feature  available   to  constrain  other  reflection
      models.  The temperature of the blackbody incident spectrum, the
      ionization parameter,  and the component  normalization are left
      free to  vary. In  this way, the  reflection model  provides the
      same information on the  ionization state of the disk reflecting
      matter, on the 1-10 keV  reflection flux, and on the hardness of
      the impinging  radiation source, as would be  derived with other
      reflection  models.  In Model  3, we  force a  spectral solution
      with disk  normalization $N$ fixed to 1200  (corresponding to an
      inner radius  of $\sim$  5 R$_{NS}$, or  $\sim$ 30  R$_g$).  For
      both  Model  2 and  Model  3,  the  smearing of  the  reflection
      component is strongly required by the fit ($\Delta \chi^2 >$ 50,
      without  any smearing), although  the smearing  parameters, when
      left free  to vary, were  strongly correlated and  uncertain. We
      chose to keep the inclination  angle and the emissivity index of
      the \texttt{rdblur} component  frozen to the best-fitting values
      of the  soft state fit (see  Table \ref{fits_1705soft}), because
      these parameters are not physically expected to vary between the
      two states.  In  Model 2 we left the inner  and outer radii free
      to   vary.  As   can  be   noted,   they  are   not  very   well
      constrained. The  inner radius  is required to  be less  than 90
      R$_g$, with a best fitting value  of 30 R$_g$. In Model 3, these
      two parameters are unconstrained as well, and we chose to freeze
      the inner radius  at a reference value of  30 R$_g$ (matching it
      with  the  inner disk  edge  value  of  the continuum  component
      \texttt{diskbb}),  while  the outer  radius  is  frozen at  1000
      R$_g$.

\begin{figure*}[ht!]
   \centering
   \caption{\scriptsize{Right panels: XMM-Newton data and residuals in
       units of  $\sigma$ with respect to the  best-fitting models for
       the hard state.  Left  panels: the theoretical unabsorbed model
       in  the 0.1--100 keV  range (in  blue the  iron line/reflection
       component,  in  green the  Comptonized  component,  in red  the
       thermal soft disk emission).}}
  \label{figmodel}
\begin{tabular}{cc}
   \includegraphics[angle=-90,width=8cm]{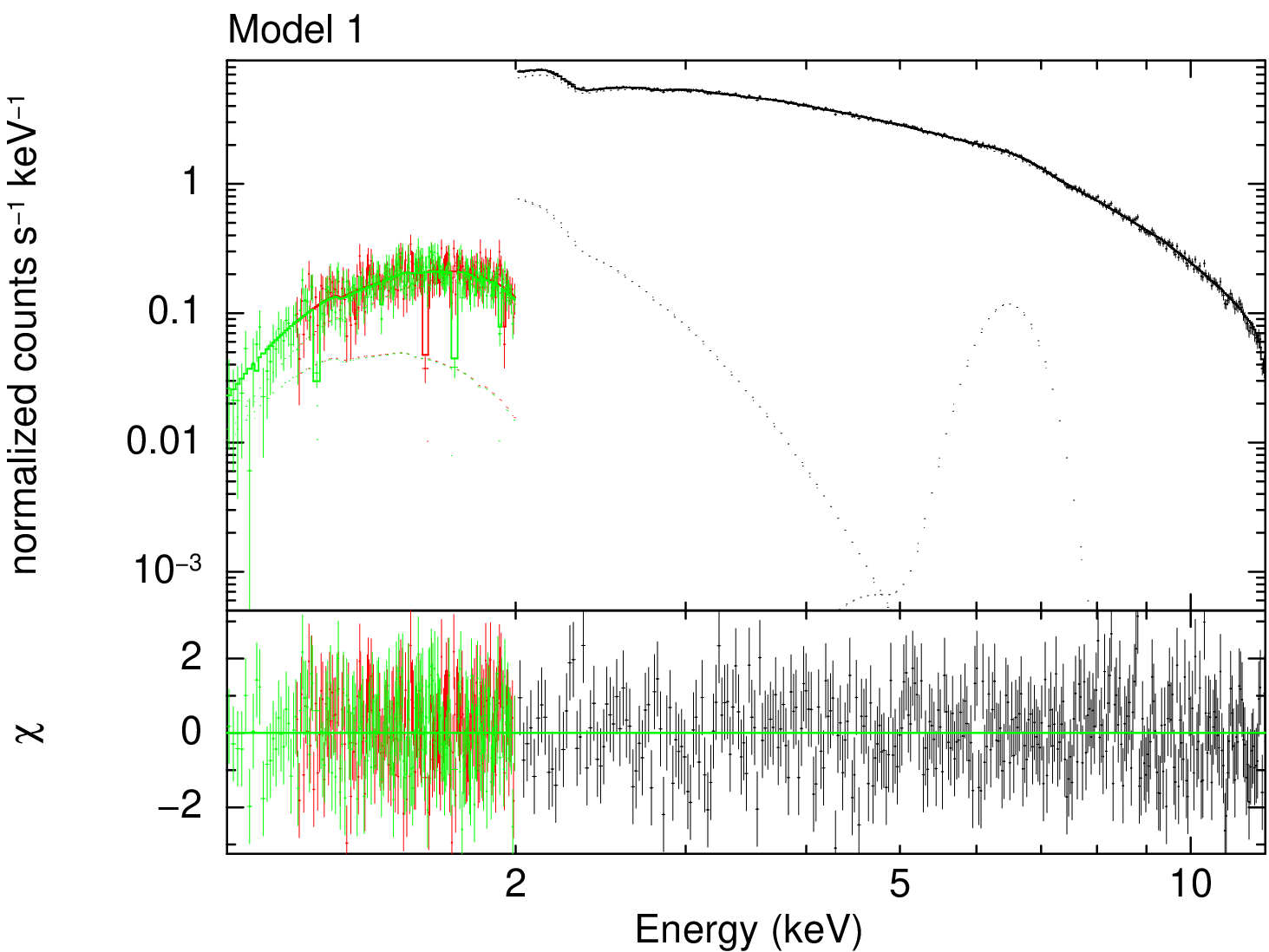}&
   \includegraphics[angle=-90,width=8cm]{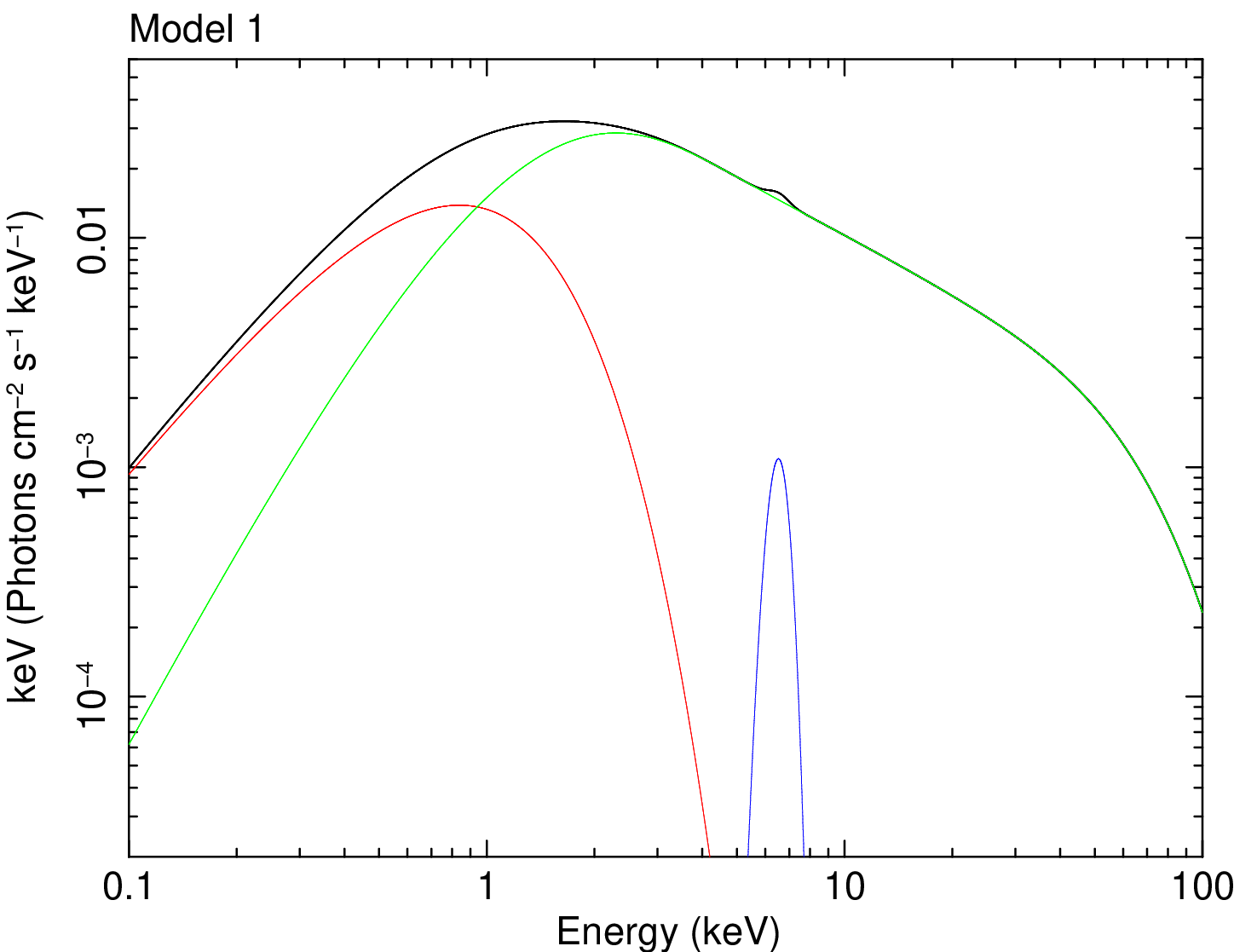}\\
   \includegraphics[angle=-90,width=8cm]{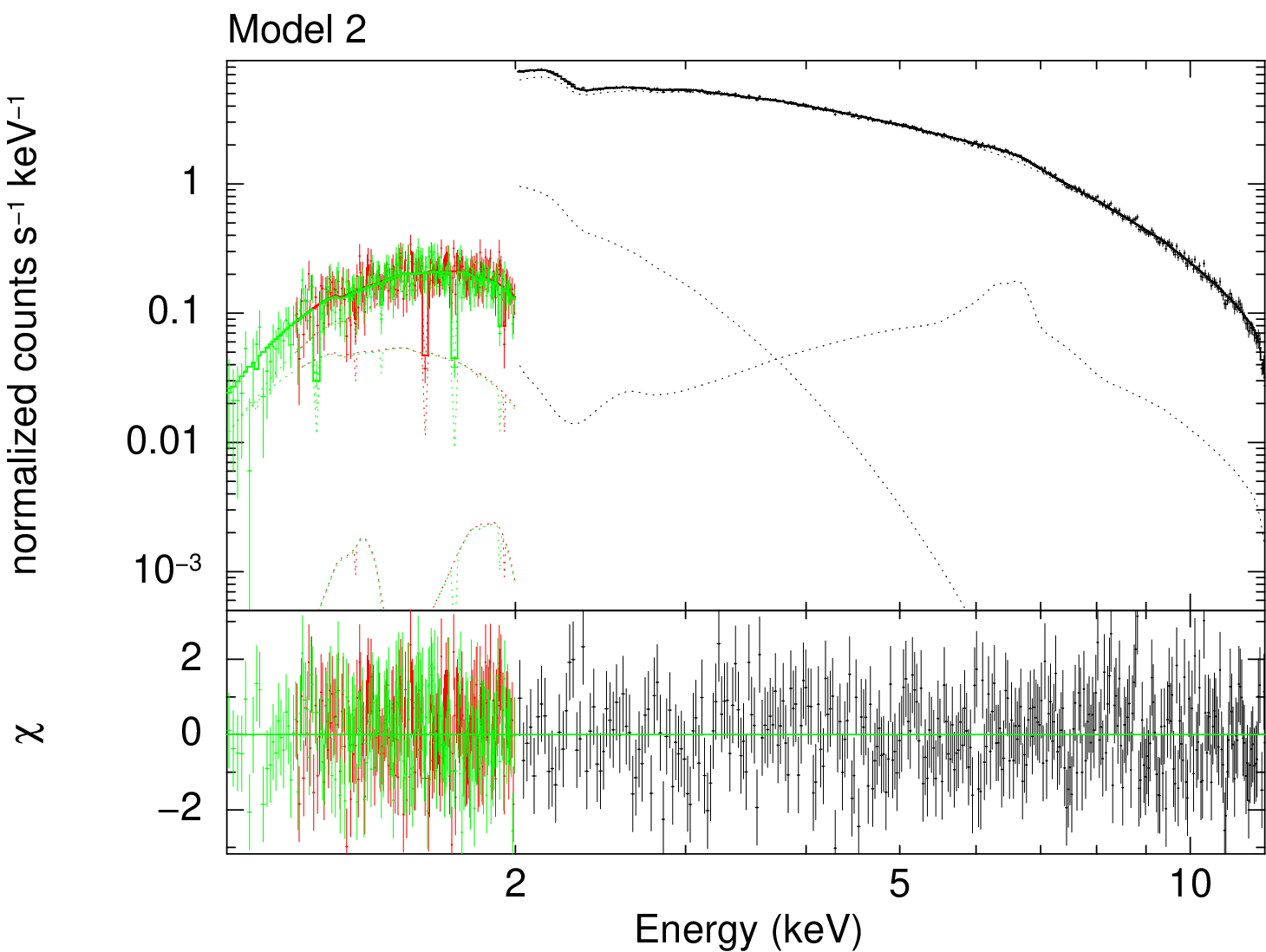}&
   \includegraphics[angle=-90,width=8cm]{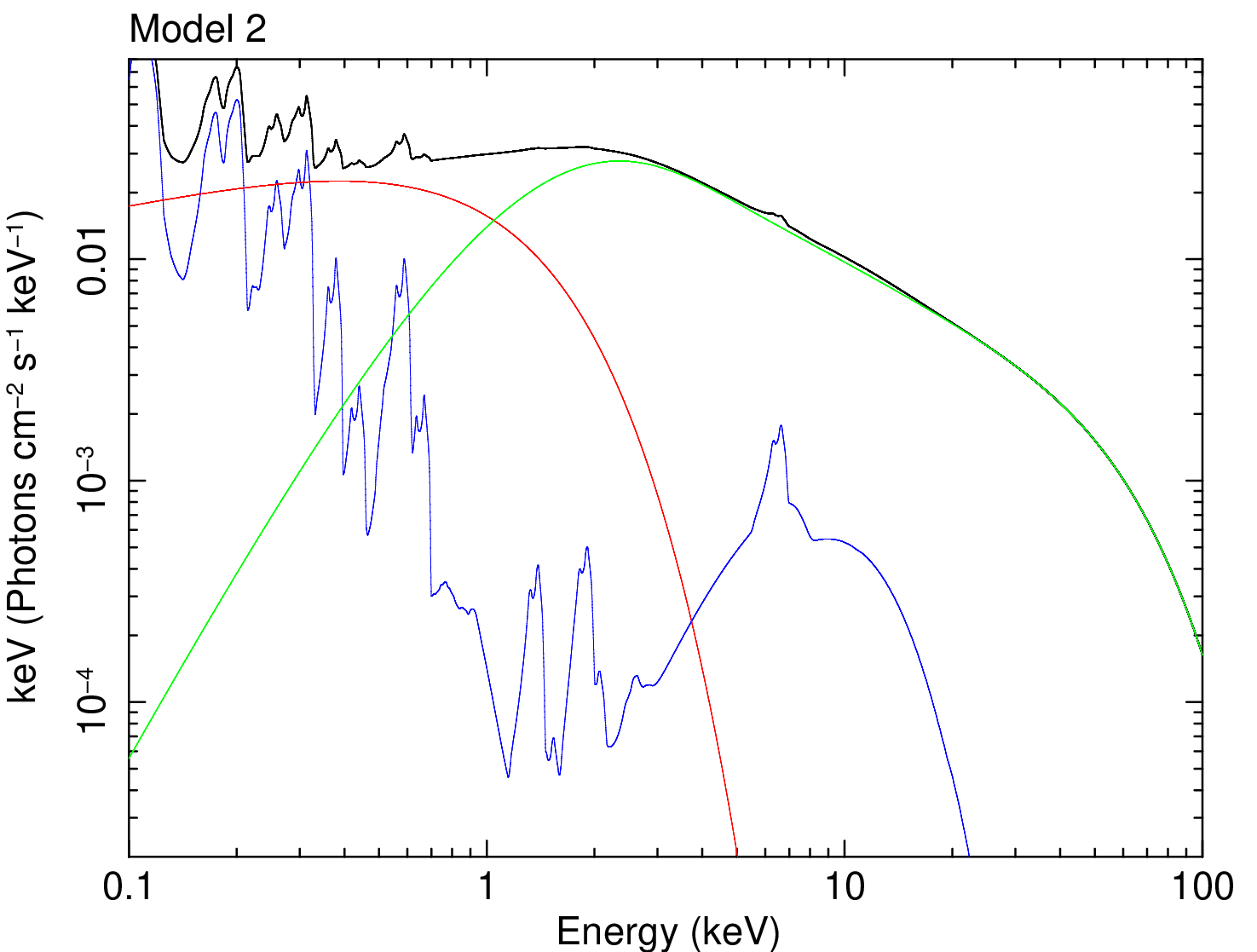}\\
   \includegraphics[angle=-90,width=8cm]{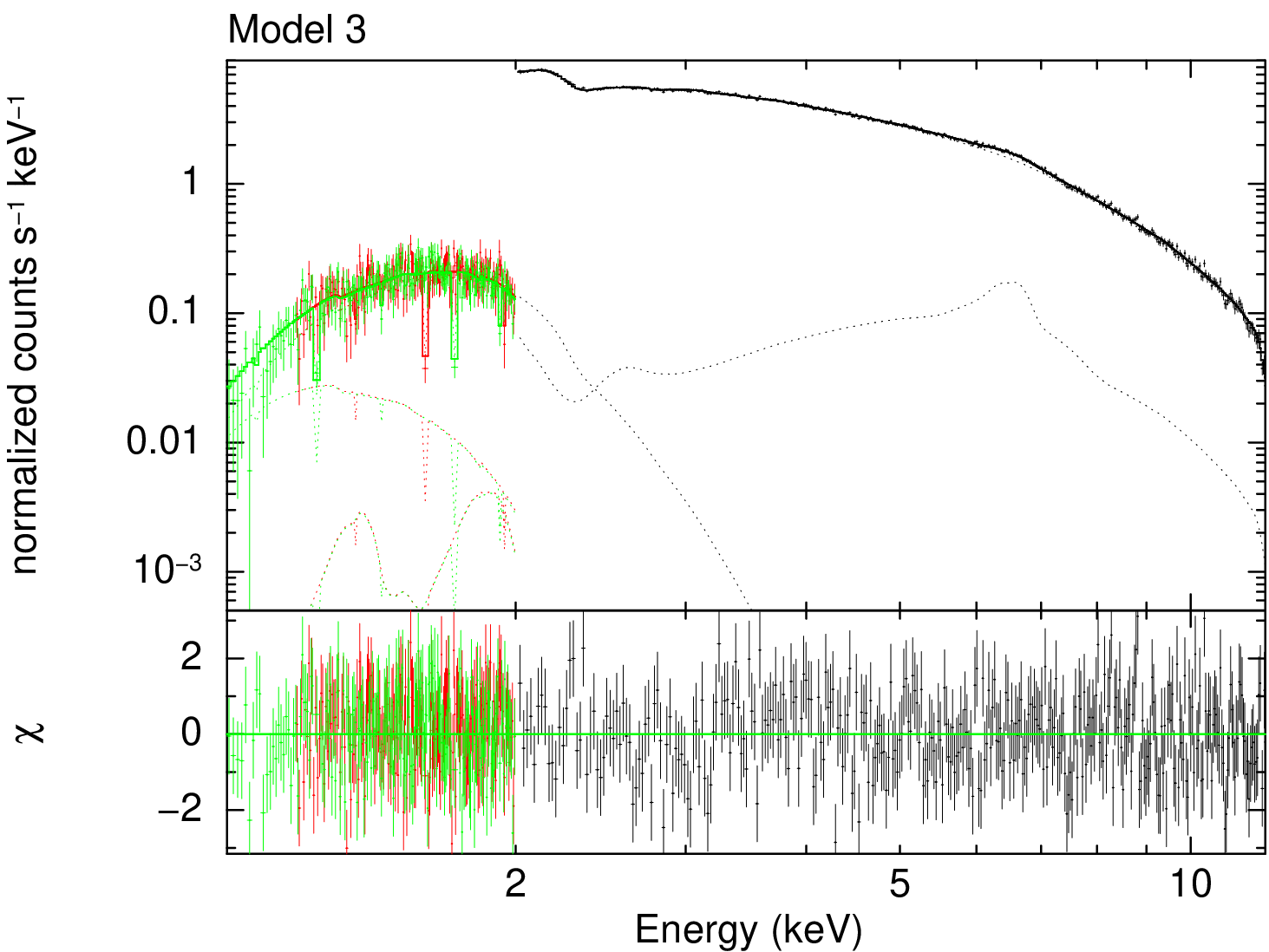}&
   \includegraphics[angle=-90,width=8cm]{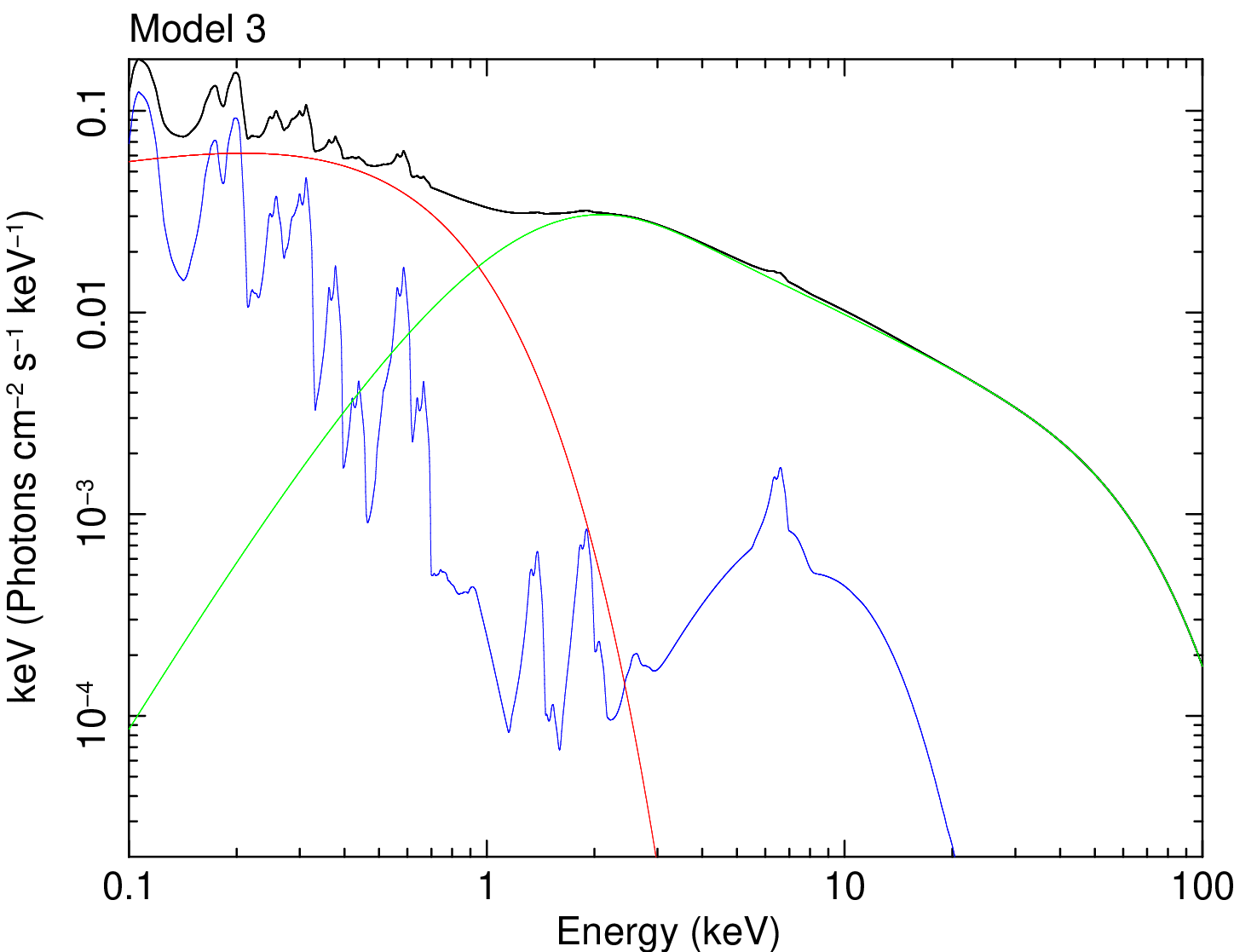}\\
\end{tabular}
    \end{figure*}

      In Fig. \ref{figmodel}  we show the data and  the residuals with
      respect   to  the   best-fitting  models,   together   with  the
      best-fitting unabsorbed model  in the extrapolated 0.1-100.0 keV
      range.   Best-fitting  values  of  the spectral  components  and
      corresponding fluxes, calculated  in the unabsorbed 1.0-10.0 keV
      range and in the  extrapolated 0.1-100.0 keV range, are reported
      in Table  \ref{fits_1705hard}. As observed,  irrespective of the
      model we  used to fit the data,  the bulk of the  emission is in
      the  Comptonized   component  and  the  fluxes   do  not  differ
      substantially from one model to the next.


\begin{table*}[ht!]
\caption{Best-fitting values and associated errors for different modelization
   used to fit the hard state data of \object{4U 1705-44}.}
\label{fits_1705hard}
\centering      
\begin{tabular}{ll lll }
\hline\hline             
Parameter              & Units                  & \multicolumn{3}{c}{Values}     \\
\hline            
                       &                    & Model 1 & Model 2 & Model 3 \\
\hline
$N_\mathrm{H}$         &  10$^{22}$ cm$^{-2}$  &   2.7$^{\rm{a}}$ & 2.7$^{\rm{a}}$ & 2.7$^{\rm{a}}$    \\  
   
Disk/BB kT        & keV            &  0.30 $\pm$ 0.04 & 0.43$\pm$0.03  & 0.23$\pm$0.12    \\
Disk Normalization &               &                         &  70$^{\rm{a}}$ &  1200$^{\rm{a}}$        \\
BB radius          & km            &  14 $\pm$ 5    &       &                  \\
Disk/BB flux      & 10$^{-10}$ erg cm$^{-2}$ s$^{-1}$ &  0.16-0.30  & 0.19-0.48     & 0.08-0.29   \\

Comptt kT$_0$  & keV    &   0.55$_{-0.02}^{+0.01}$ & 0.57 $\pm$ 0.03  &   0.50 $\pm$ 0.01 \\
Comptt kT$_e$  & keV    &   14.4 $\pm$ 0.2        & 13.8$_{-0.3}^{+0.2}$  & 13.8$_{-0.3}^{+0.2}$ \\
Comptt $\tau$  &        &   5.5$^{\rm{a}}$  & 5.5$^{\rm{a}}$ & 5.5$^{\rm{a}}$  \\
Comptt flux    & 10$^{-10}$ erg cm$^{-2}$ s$^{-1}$ &   2.6-6.2   & 2.5-5.7  & 2.1-5.2   \\

Iron line      & keV                              & 6.50 $\pm$ 0.07 &            &       \\
Iron $\sigma$  &                                  & 0.41 $\pm$ 0.08 &            &        \\
Line EW        & eV                               & 60 $\pm$ 25                  &         \\

Inner radius   & R$_g$                            &  & 30$_{-24}^{+60}$   &  30$^{\rm{a}}$                  \\
Outer radius   & R$_g$                            &  & 700$_{-300}^{+5000}$ &  1000$^{\rm{a}}$       \\
Betor          &                                  &  & 2.3$^{\rm{a}}$ &  2.3$^{\rm{a}}$              \\
Inclination    & degrees                          &  & 35$^{\rm{a}}$  & 35$^{\rm{a}}$      \\

$kT_{ion}$     & keV     &               & 1.8 $\pm$ 0.3 & 1.6 $\pm$ 0.2                 \\
log $\xi$      & &                       & $<$ 1.1      & $<$ 1.04                      \\

Refbb Flux           & 10$^{-10}$ erg cm$^{-2}$ s$^{-1}$ & & 0.08-0.24   &  0.08-0.29 \\

$\chi^2_{red}$ (d.o.f.)                            & & 743 (727)  & 747 (726)  & 753 (726) \\
\hline
\multicolumn{5}{l}{$^{\rm a}$  Frozen parameter during the fitting procedure}\\
\end{tabular}
\end{table*}


\section{Burst analysis} \label{burstanalysis}

In HardObs,  the source showed a type  I X-ray burst. The  peak of the
burst was observed 10.844 s after the beginning of the observation and
lasted approximately  for 100  s. The light  curve profile  during the
decay phase is  well-fitted by an exponential curve  with a decay time
$\tau$ of 10.4  s.  The spectral evolution of the  burst is similar to
what is observed  in other sources.  We used the  source spectrum 15 s
before  the  onset  of the  burst  as  background  file, and  then  we
extracted a source spectrum every 5 s, after the burst peak. The burst
spectrum is well-fitted by an absorbed blackbody spectrum and
does  not  show  any other  spectral  feature,  also because  the  low
statistics and the short  exposure times.  In Table \ref{table_burst},
we show the variation in the burst flux and the decreasing temperature
of the continuum component as a  function of time.  The burst does not
show photospheric radius expansion.

\begin{table}[h!]
\caption{Time-resolved spectral evolution of the burst emission.}
\label{table_burst}
\centering      
\begin{tabular}{llll}
\hline\hline             
        & BB kT & BB luminosity & BB radius \\
\hline
Units   & keV   & erg s$^{-1}$ ($\times$ 10$^{37}$) & km \\
\hline
Time 1  & 2.91$\pm$0.15    &  15.8$\pm$1.5         &  4.1$\pm$0.6  \\  
Time 2  & 2.41$\pm$0.11    &  10.2$\pm$0.8         &  4.8$\pm$0.6  \\ 
Time 3  & 1.74$\pm$0.07    &   4.0$\pm$0.2         &  5.8$\pm$0.7  \\ 
Time 4  & 1.53$\pm$0.09    &   2.0$\pm$0.1         &  5.4$\pm$0.8  \\ 
Time 5  & 1.36$\pm$0.09    &   1.4$\pm$0.1         &  5.7$\pm$0.9  \\ 
Time 6  & 1.26$\pm$0.09    &   0.96$\pm$0.08       &  5.4$\pm$1.0  \\ 
\hline                               
\end{tabular}
\end{table}

\begin{figure}[ht!]
   \centering
   \resizebox{\hsize}{!}{\includegraphics[angle=-90]{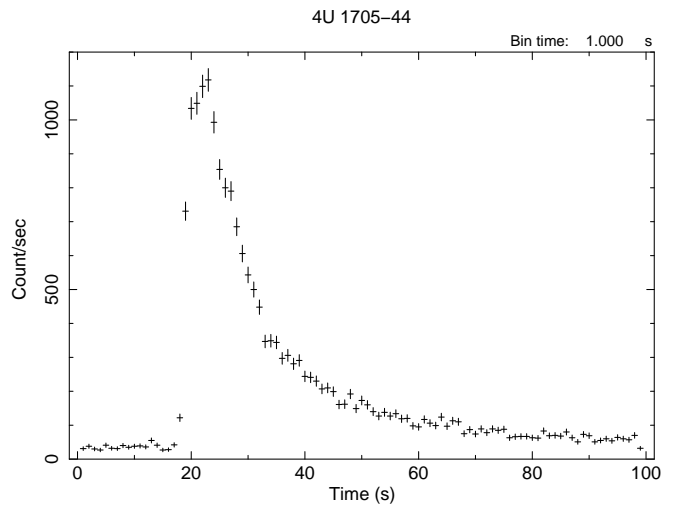}}
   \caption{\scriptsize{Light  curve of the burst observed
       during HardObs, the start time  of the first time interval used
       for spectral analysis is time=20  s in this plot.}}
              \label{lc_burst}%
    \end{figure}

\begin{figure}[h!]
   \centering
   \resizebox{\hsize}{!}{\includegraphics[angle=-90]{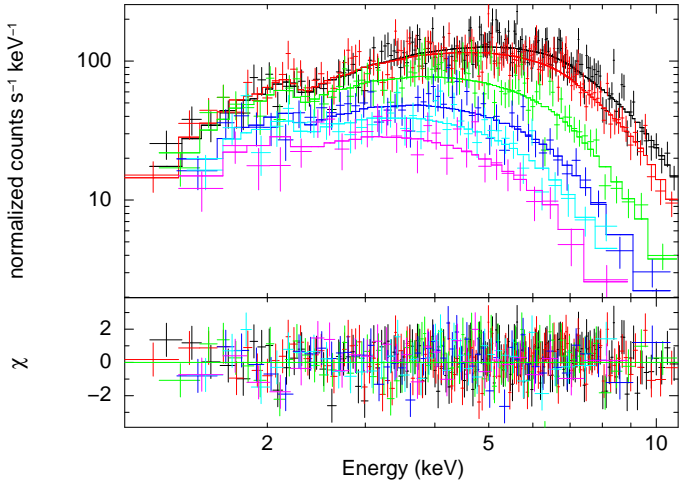}}
   \caption{\scriptsize{
       Data and residual with respect to the best-fitting model of the
       six time selected spectra of Table \ref{table_burst}.}}
              \label{lc_burst2}%
    \end{figure}

The study of this burst allows,  for the first time in this source, to
independently  constrain  the  value  of the  interstellar  absorption
towards the  source direction.  Past  studies, in fact, relied  on the
Rossi XTE data, where the spectra started from 3.0 keV and gave nearly
no constraint on the $N_\mathrm{H}$.  We fitted the six datasets using
a  common fit,  with the  $N_\mathrm{H}$ tied  for each  dataset.  The
reduced $\chi^2$  of the fit is  acceptable, 1.03 for 472  dof, and we
derived  a  value for  the  $N_\mathrm{H}$  parameter =  2.46$\pm$0.22
$\times$   10$^{22}$  cm$^{-2}$   (2.46$\pm$0.35   $\times$  10$^{22}$
cm$^{-2}$  for   99\%  confidence  level).   Using   the  once  common
\texttt{wabs} interstellar absorption component, we get $N_\mathrm{H}$
= 1.70$\pm$0.15 $\times$ 10$^{22}$  cm$^{-2}$. We further checked that
this global $N_\mathrm{H}$  value is within the error  bars, when this
parameter is independently  free to vary for each  spectrum. We report
in  Table  \ref{table_burst}  the  best-fitting  blackbody  parameters
(blackbody radius calculated  assuming a distance of 7.4  kpc) for the
six time intervals and in Figs.  \ref{lc_burst} and \ref{lc_burst2} the
burst light curve and the data with the residuals in units of $\sigma$
for the common fit.

The evolution of this isolated burst  is similar to some of the bursts
reported for this source in the RXTE burst catalog \citep{galloway08}.

\newpage
\section{Discussion}

\subsection{The soft state}

The two XMM-Newton observations of \object{4U 1705-44} offer the opportunity of
testing a self-consistent model for  the 1.0-12.0 keV energy band and,
using spectral fitting, probing the different accretion states in this
source. During  SoftObs the source  was in a high-flux, soft
state.  The bolometric luminosity of  the source was at $\sim$ 50\% of
the  Eddington limit,  with  most  of the  X-ray  emission within  the
observed 1-12 keV band ($\sim$ 70\%). This state corresponds to one of
the softest states ever observed  for this source. Because most of the
flux, theoretically limited by the Eddington rate, is radiated at soft
energies (E$<$10 keV), we expect  that the spectrum is relatively weak
in  hard  X-rays,  \citep{barret02}.   Even considering  the  possible
presence of a  nonthermal hard tail, which was reported when the
source luminosity  corresponded to  4 $\times$ 10$^{37}$  erg s$^{-1}$
\citep{piraino07},  this characterization  remains valid  as  the hard
tail flux amounted to only $\sim$ 10\% of the total source luminosity.

 The spectrum can be well-fitted  by the sum of two thermal components
 together with  the reflection component. This model  accounts for the
 energetics and spectral  distribution between the different continuum
 components,  if we  associate the  blackbody hard  emission  with the
 boundary    layer    (BL)    emission   \citep{done02,    gilfanov03,
   revnivtsev06}.  The  boundary layer should, in fact,  occupy a very
 small region between the inner disk  radius and the surface of the NS,
 and theoretically,  it should Comptonize seed photons  from the disk
 and  the surface  of the  NS in  an optically  thick corona  in these
 high-accretion  states  \citep{popham01},  resulting in  a  saturated
 single-temperature Comptonization.

 The blackbody radius of the hard component is 5.5 km, a value that is
 remarkably  in agreement with  the asymptotic  value that  we derived
 from the analysis of the burst emission.  Taking general relativistic
 effects   into   account   and   the  hardening   of   the   spectrum
 \citep{shimura95}, this radius is  compatible with an emission region
 of  size comparable  to the  NS  surface. The  disk and  the BL  have
 comparable fluxes  with a  luminosity ratio $L_{BL}/L_{disk}$  = 1.67
 (assuming a  spherical isotropic emission  for the BL  and neglecting
 the  reflection contribution)  or  2.0 (considering  the  sum of  the
 reflection emission and the hard component).

We can express the ratio of these two components in terms of the outer
boundary layer radius, which can be  written in units of the NS radius
as  $b  R_{NS}$, and of the  spin frequency of the NS,  which can be
written  as  a fraction  of  the  breakup  spin frequency,  $f  \equiv
\nu_{spin}/\nu_{break}$ \citep{popham01},

\begin{equation}
  \frac{L_{BL}}{L_{disk}} = \frac{f^2}{b} - 
  2\phantom{.}b\phantom{.}\sqrt{b}\phantom{.}f + 2\phantom{.} b - 1.
\end{equation}

The spin  frequency of  the NS is  not known; however,  the difference
between the upper and the lower kHz QPOs is, in some sources, near the
true  spin   frequency  \citep[see][]{vanderklis04,mendez07}.   In  4U
1705-44,  this  difference  is  $\sim$ 330  Hz  \citep{ford98},  which
translates into a $f$ value between 0.13-0.2 depending on the equation
of state adopted for the NS. A ratio of 2, as we found, constrains the
$b$ value  in a very  narrow range between  1.8 and 2.1,  which agrees
with theoretical expectations  for such a high accretion  rate, of the
outer radius of the  boundary layer \citep{popham01}.  In some sources
the difference  in kHzQPOs peaks is  near half the  spin frequency. If
this were the case, on the  contrary, we would obtain $b$ values above
2.6, contradicting both the theory  and the inner disk radius obtained
from the reflection component.

The thermal disk  component normalization $N$ is related  to the inner
disk  radius by  the  simple expression  $R_{col}  = D  \sqrt[2]{N/cos
  \theta}$, where $D$ is the distance expressed in units of 10 kpc and
$\theta$ is the inclination angle.  Using a distance of 7.4 kpc and an
inclination angle  of 35$^{\circ}$, we obtain a  best-fitting value of
11.4   km.    However,  this   value   is   generally  considered   an
underestimation  of the  true inner  radius by  a factor  of  $\sim$ 2
\citep{merloni00}.  Taking  these corrective factors  into account, we
derive an  inner disk radius of  $\sim$ 10 R$_g$,  consistent with the
smearing factor of  the reflection component and with  our estimate of
the  outer  boundary layer  radius  as derived  by  the  ratio of  the
continuum fluxes \citep{dai07,dai09}.

The reflection component  in this state is remarkably  strong, and the
reprocessed  flux at  the  disk surface  is  about 30\%  of the  total
Comptonized component.  Although most  clearly determined by the shape
of the broad iron line,  the reflection component, gives its strongest
contribution in the very soft X-ray band (below 1 keV there is 62\% of
the overall bolometric flux of the component, with the dominant H-like
lines of oxygen and neon), where  it can be as competitive as the disk
emission, as  already noted by \citet{ballantyne04}.  In  the 1-10 keV
range, the relative contribution is 33\%, while above 10 keV, there is
only  a 5\%  flux,  owing to  the  very soft  nature  of the  incident
flux. We calculated the reflection fraction ($R = \Omega / 2 \pi$) for
this model to be 0.23 $\pm$  0.01, a value that is consistent with our
proposed geometry of a compact illuminating source.

The disk  approaches the NS,  and we obtain accurate  constraints from
the  relativistically  broadened reflection  features,  which set  the
inner  disk  radius  to  less  than 13  R$_g$.   After  comparing  our
best-fitting model with the results of \citet{disalvo09}, we note that
the blurring  parameters agree quite  well, with a difference  only in
the  outer radius,  which in  our  model is  found larger  and in  the
inclination angle,  which is required  to be somewhat  lower (although
still  compatible  at  3  $\sigma$  level).   These  differences  are,
however,  impressively small  if we  consider that  we adopted  here a
different continuum  model and a  self-consistent broadband reflection
model instead of local disklines.


\subsection{The hard state}

Bright Z-sources accreting  very close to the Eddington  limit, do not
show significant  variation in the  luminosity ratio between  disk and
coronal emission when  they move along their Z-track  on a color-color
diagram,  and the  evolution  of the  spectral  parameters is  smooth,
without jumps \citep[e.g.][]{disalvo02,agrawal03,dai07}.

Atoll  sources, in  contrast, resemble  the spectral  decomposition of
bright  Z-sources when  they  are  in their  soft  state (also  called
\textsl{banana state}),  but strongly differ  when they move  into the
hard state (also called \textsl{island  state}).  In this state a very
soft  component is  occasionally  detected \citep{barret00,disalvo04},
although  the inferred  emission  radii are  of  a few  kms, and  some
authors identify  this soft emission  directly with the  NS surface/BL
emission  \citep{gierlinski02,barret03}.  The luminosity  is dominated
by a Comptonized  spectrum, with moderate optical depth  (1 $<$ $\tau$
$<$ 6) and  high-energy cut-offs between 30 and 100  keV.  There is no
general agreement about what  causes these state transitions, although
it is  certain that  the soft-hard and  the hard-soft  transitions are
switched at different luminosities giving rise to a hysteresis pattern
\citep{meyer05}.   The difference  in  the switch  luminosity is  not,
however,  universal,  and  4U  1705-44  displays  a  small,  or  null,
hysteresis \citep{gladstone07}.  The  luminosity that we derived using
the models 1-3 is within 2-4 $\times$ 10$^{36}$ erg s$^{-1}$, which is
slightly below  the threshold that \citet{gladstone07}  quotes for the
state transition of \object{4U 1705-44}.

Comparing these results  with the models proposed for  the soft state,
we  note the  very small  contribution of  the soft  component  to the
overall energy emission.  The source  accretes at levels less than 1\%
L$_{Edd}$,  so that  the  disk is  theoretically  expected to  closely
approach the NS  surface (at 1.1-1.2 R$_{NS}$), because  it has a flux
contribution  higher,   or  similar,  to   the  Comptonized  component
\citep{popham01}.  But,  as can be seen  in Table \ref{fits_1705hard},
any soft  component would make only  $\sim$ 1/10 of  the overall total
contribution. This aspect is well known as are the difficulties in the
spectral decomposition \citep[see][for a complete discussion.]{lin07}.

  We showed the  best-fitting parameters in Table \ref{fits_1705hard},
  and the  residuals and the  extrapolated models for  three different
  modelizations,  which  are able  to  describe  the  1-10 keV  energy
  spectrum in Fig. \ref{figmodel}.  The model with the lowest $\chi^2$
  is  Model 1,  although  the  two other  models  cannot be  \textsl{a
    priori} ruled out, as the difference in $\chi²$ is not significant
  and is given  to tighter constraints imposed on Models  2 and 3.  We
  discuss  hereafter the  physical implications  of the  three models,
  focusing  on the  combined broadband  continuum and  line broadening
  mechanisms.

\subsubsection{Model 1}

This model provides a very good  fit to the data, and is also proposed
in \citet{barret02} for RXTE spectral  analysis of the same source and
in \citet{gierlinski02}  for the atoll source 4U  1608-52.  It closely
resembles       the       so-called      \emph{Birmingham       Model}
\citep{church01,church04},   where  the   spectral   decomposition  is
resolved  into a  soft blackbody  emission, ascribed  to  the boundary
layer,  and a  Comptonized emission  that takes  place in  an extended
corona (coronal radius  $>> 10^9$ cm) caused by  disk evaporation. The
inferred blackbody radius is rather small  (14 $\pm$ 5 km), but it can
be  still consistent  with a  surface physically  associated  with the
BL. We  derive good constraints  for the Comptonized component  on the
seed photon  temperature and the  thermal temperature of  the electron
cloud. When  comparing our  results with \citet{barret02},  we observe
that the  soft blackbody component  has a much lower  temperature (0.3
keV against 1 keV), while  the electron temperature is well consistent
with  the  RXTE  results.   The discrepancy  is,  however,  explicable
considering  the  much better  low  energy  coverage  of the  spectrum
offered by XMM-Newton, while the RXTE  data start only from 3 keV, not
allowing  any distinction  between  the Comptonized  curvature at  the
seed-photon temperature and the contribution of any other soft thermal
component.  (In \citealt{barret02}, the seed-photon temperature is, in
fact,  kept  frozen  to a  reference  value  of  0.4 keV  because  not
constrained by the fit.)

We  note, however, that  the blackbody  temperature is  too low  to be
associated with the reprocessed emission from the NS surface, while it
is physically implausible to  observe seed photons for the Comptonized
component more energetic than the emitted photons for the BL or the NS
surface.  We find  it more reasonable, on the basis  of the results of
the soft  state, to consider  that the continuum soft  component still
needs  to be  associated with  the  thermal disk  emission, while  the
Comptonized emission continues between the inner edge of the accretion
disk and the surface of the NS.

\subsubsection{Model 2}

In analogy with  soft state, this model ascribes  the softer continuum
component  to  thermal  disk  emission  and the  harder  component  to
Comptonized emission.   We tried two  spectral solutions to  check, at
least as  a first  order approximation, that  the iron  line broadness
could still be  interpreted in terms of disk  reflection. An optically
thick,  geometrically thin  disk must,  therefore, be  present,  so we
tried  in  Model 2  to  fit  the low-energy  X-ray  band  with a  disk
emission,  whose  inner radius  would  match  an  R$_{in}$ $\sim$  1.2
R$_{NS}$,  as   is  theoretically  expected   \citep{popham01}.   This
corresponds  to  fixing   the  normalization  of  the  \texttt{diskbb}
component to 70, a value we found  to lie within the error box, if the
parameter is  left free to  vary. The reflection component  requires a
$kT_{ion}$ temperature between 1.5 keV and 2.1 keV.

The  degree of  smearing,  given  by the  constraints  imposed on  the
\texttt{rdblur}  component are sufficient  to adequately  describe the
iron  line  broadness.  In  this  case,  the  reflection has  a  lower
subtended angle with  respect to SoftObs, and the  percentage ratio of
the  extrapolated  fluxes  between   the  corona  and  the  reflection
component  is  $\sim$  4\%,  which  can  physically  correspond  to  a
geometrical  shrinking   of  the  area   of  the  BL,  which   from  a
quasi-spherical shape around the NS  in the soft state, is expected to
decrease its area  to an equatorial belt of  a few kilometers diameter
\citep{inogamov99}.   Half of  the  reflected energy  is absorbed  and
re-emitted at energies less than 1 keV, while a higher fraction, about
20\%, with respect to the soft state, emerges above 10 keV.

This  modelization has,  however, the  main flaw  of  unexplaining the
strong difference  in the  fluxes of the  two components (more  than a
factor of 10).  A similar  spectral decomposition is also possible for
black  hole  sources in  their  low/hard  states,  where a  very  soft
component,  together  with  a   reflection  component  arising  in  an
untruncated accretion disk,  is claimed to be present  by some authors
\citep{miller06a,miller06b,rykoff07},  but it  is  rejected by  others
\citep{done06,        dangelo08,       gierlinski08,       hiemstra09,
  cabanac09}. However, in black holes, there is no balance required in
the  energetics of  the  components  that make  up  the overall  X-ray
emission, as the quantity of  energy advected beyond the event horizon
is  an unknown  variable. In  contrast, in  X-ray systems  harboring a
neutron star, all the accretion energy must be dissipated and released
as  visible electromagnetic  radiation (excluding  any  strong beaming
effect that could  arise in a state, but not  in an another).  Because
the  spectral solution  that we  envisioned for  SoftObs  is perfectly
consistent with the presence of a standard accretion disk truncated at
$\sim$ 2 R$_{NS}$, from an energetic distribution between the spectral
components, we conclude that, although statistically consistent, Model
2  lacks   a  correspondingly  coherent  physical   scenario.  It  is,
therefore, possible that the  disk structure strongly departs from the
standard   Shakura-Sunyaev    $\alpha$-disk   model,   becoming   less
energetically  efficient or  losing  its thinness  \citep{maccarone03,
  meyer-hofmeister05, dullemond05}.

Recently,  \citet{taam08}  explored  this  scenario for  the  low/hard
states of black  holes, finding that at luminosities  of a few percent
of  the Eddington  limit,  a cool,  underfed,  disk can  be formed  by
condensation  of coronal  matter onto  the  disk. This  disk would  be
under-luminous with respect to  a standard $\alpha$-disk and still act
as a reflecting medium for coronal photons.  However, the shape of the
reflection    component    would    be    strongly    influenced    by
Compton-scattering  in the corona,  if its  optical thickness  is much
higher than 1,  as observed in the hard state of  this source, and the
line   should  be  strongly   smeared  and   practically  undetectable
\citep[see the  discussion in][]{disalvo05}.  We are therefore  led to
consider  a  different  scenario,  where  still  we  have  a  standard
accretion disk, but  truncated at some distance from  the neutron star
by an unknown physical switch (magnetic pressure of the B-field of the
NS or change  in the accretion flow or evaporation  or a still unknown
physical mechanism).
 
\subsubsection{Model 3}

This model revisits Model 2  by forcing a spectral solution consisting
of a  truncated accretion  disk, Comptonized boundary  layer emission,
and  reflection component. The  model is  adequate for  describing the
shape  of the  line,  and  it is  only  marginally statistically  less
favored than  the two other models.  The \texttt{diskbb} normalization
is set  to correspond to  a truncated disk  at 30 R$_g$ and  the inner
radius  of  the  reflection  component  is also  frozen  at  the  same
value. We  are not interested,  here, in evaluating where  exactly the
disk  is  truncated, as  we  do  not  have the  necessary  statistical
constraints  for  it, but  rather,  we  are  interested in  showing  a
representative  spectral solution, which  is physically  motivated and
still statistically acceptable.

Constraining the  normalization of the \texttt{diskbb}  component to a
higher value  than for Model 2  produces in the best-fit  a lower disk
temperature (from 0.4 keV to  0.2 keV) and a lower Compton seed-photon
temperature while  it has no effect  on the determination  of the hard
part of the spectrum and on the shape of the reflection component.

We  observe  a  large  change  in  the  ionization  parameter  of  the
irradiated disk when  we compare the reflection component  in the hard
and soft states. The $\xi$ parameter, defined:
\begin{equation}
\xi = \frac{4 \pi F_x}{n_H}
\end{equation} 
can be  used to  compare the changes  in the impinging  flux, assuming
that  density  and  distance  do   not  strongly  differ  in  the  two
observations.

From $\xi_1 /  \xi_2 \simeq F_{1x}/F_{2x} $ and  the results quoted in
Tables \ref{fits_1705soft} and  \ref{fits_1705hard}, we would expect a
change in  the fluxes of an  order of magnitude,  which corresponds to
the flux ratio  of the hard blackbody component of  the soft state and
the  Comptonized flux of  HardObs. We  also calculated  the reflection
fraction  for this hard  state.  However,  because the  hard continuum
component in this  state is a thermal Comptonized  component and not a
blackbody, we measured the  reflection fraction only in the restricted
energy band (5.5-8 keV),  where we approximated the continuum emission
with a  blackbody-like spectrum. We  obtained a value  consistent with
the reflection fraction in the soft  state (0.2 $\pm$ 0.1), but with a
larger  uncertainty. This value  indicates that  the geometry  has not
noticeably  changed and that  the scenario  of a  compact illuminating
source is still valid.

\subsection{The reflection scenario in the hard and soft states}

The analysis  of the XMM-Newton data  in the soft state  of \object{4U 1705-44}
presented  in this  work  confirms the  results of  \citet{disalvo09}
regarding the  reflection origin of the local  emission and absorption
features and their  relativistic broadening.  Our re-analysis excludes
any possible  influence of  pile-up on the  spectral modeling  of the
continuum and of  the broad emission lines. Still,  the quality of the
data  allows for tight  constraints on the main  spectral parameters
that are at the origin of the broadening.
 
In  \citet{disalvo09}, the  spectrum  shows also  a  broad feature  in
absorption  seen  at  8-8.5  keV,  which has  been  interpreted  as  a
gravitationally  red-shifted  smeared  K$\alpha$  absorption  edge  of
He-like  iron ions.   The presence  and the  degree of  this smearing,
however, becomes  very significant  only when the  central row  of the
Epic-pn  CCD is  considered.   In our  re-analysis  (that reduces  the
statistics  of a  factor of  2 with  respect to  a  spectrum extracted
including  the central  row), the  absorption  edge is  detected at  a
weaker confidence  ($\Delta \chi^2$ =  -30, with 3 d.o.f.   more, than
for the  same spectral model of \citet{disalvo09}  without the smeared
edge), but the best-fitting spectral parameters remain consistent with
the values reported in \citet{disalvo09}.  In our re-analysis both the
absorption and the emission  processes are consistently described by a
smeared broadband  reflection model that has  a blackbody prescription
for the irradiating source photons.
  
Suzaku  data  of  4U  1705-44  also support  the  reflection  scenario
interpretation  for   the  observed  broad   features  \citep{reis09}.
Although  the spectral  decomposition is  different, both  the present
work and  the analysis  of the Suzaku  data clearly indicate  that the
reflection component can vary according  to the accretion state of the
source, that the  source is seen at an  intermediate inclination angle
(30  $\pm$  1 deg  in  the  Suzaku  fits), that  relativistic  effects
dominate because they arise from  an accretion disk extremely close to
the  compact  object (R$_{in}$  =  10.5$_{+1.0}^{-1.7}$  R$_g$ in  the
Suzaku fits),  and that the  ionization degree of the  disk reflecting
matter strongly depends  from the accretion state of  the source (with
ionization  parameter values  ranging  from 10  to  3500 erg  cm$^{2}$
s$^{-1}$ in the  Suzaku fits).  The model of  the present paper shows,
however, some important differences  with respect to the model adopted
in \citet{reis09}:  the continuum thermal disk  emission is consistent
with  a thermally hot  accretion disk,  close to  the NS  surface; the
reflection model  has a different prescription  regarding the spectral
shape of the disk illuminating  primary source; i.e., we propose that,
during  the  soft  state,   the  Comptonized  component  approaches  a
blackbody-like  spectrum  and  this  is  the  source  of  the  primary
impinging  radiation  on  the  disk, whereas  in  \citet{reis09},  the
\texttt{reflionx}  table  model \citep{ross05}  has  been used,  which
assumes a power-law spectral distribution of the irradiating flux.  We
showed that the interpretation of  a thermalized source for the disk's
illuminating  source   is  justified  both  by  theory   and  by  good
model-to-data fitting for the soft state.

This  assumption  may  not  be  true  for  hard  states,  because  the
Comptonized component  is no longer optically thick,  and the spectrum
substantially hardens, with most of  its flux above 10 keV. Therefore,
to  test both  the  continuum model  and  the shape  of the  reflected
component, a  coverage of  the hard X-ray  spectrum (above 10  keV) is
essential.  If the  nature of the disk's impinging  radiation is still
quite soft, with a low ionizing temperature (less than 3 keV), then no
reflection Compton  curvature is expected at higher  energies, while a
Compton  bump  should  be  present,  together  with  the  fluorescence
emission  lines,  in  the  case  of an  extended  power-law  impinging
spectrum.  A claim of a detected reflection Compton bump has been made
by \citet{fiocchi07} using INTEGRAL data; however, the strength of the
Compton  bump was not  correlated with  the iron  line flux,  and this
would contradict a common physical origin.


\section{Conclusions}

We have examined and compared two XMM-Newton observations of the Atoll
source 4U  1705-44, in a soft and  in a hard spectral  state. The soft
state  is   characterized  by   three  main  spectral   components:  a
multicolored thermal disk emission, a harder, blackbody-like, boundary
layer emission, and a relativistically smeared reflection component. A
self-consistent model  of reflection has been  successfully applied to
fit  the  data, where  the  incident spectrum  is  assumed  to have  a
blackbody shape.  We have shown  that the temperature of  this thermal
irradiating  flux  is  consistent  with  the  thermal  boundary  layer
temperature and found from independent constraints that the inner disk
radius, coincident with the outer  boundary layer radius is located at
a distance of  $\sim$ 2 R$_{NS}$.  The second  observation, taken when
the source  is in  the hard  state, is also  characterized by  a broad
emission  feature in  the  iron range,  although  its physical  origin
appears less constrained because of the much lower statistics and lack
of high-energy response.  We applied different modelizations, and with
the constraints obtained  in the analysis of the  soft observation, we
showed  and  discussed three  possible  spectral decompositions.   The
scenario  that appears  more successful  in  data fitting  and in  its
physical implications is  a model of very soft  thermal disk emission,
with an inner edge truncated  at relatively large distance from the NS
(between 20  R$_g$ and 90  R$_g$) and a thermal  Comptonized emission.
The  width of  the iron  line in  this state,  can still  be explained
within  the  reflection  scenario,  even with  a  truncated  accretion
disk. Although the  line is intrinsically broad, it  does not show any
evidence of asymmetry, because of the lower statistics with respect to
the  soft  state  observation  and because  relativistic  effects  are
strongly reduced if  the disk is truncated at  a greater distance from
the NS. To address the nature  of the reflection component in the hard
states,  we still  need good  spectral coverage  of the  overall X-ray
emission, in the soft (less than  1 keV) range to better constrain the
continuum disk emission, in the iron range to resolve the shape of the
iron  line, and  in the  hard (above  10 keV)  range to  constrain the
reflection    Compton-scattered   continuum.    

\vspace{0.5cm}

During the peer review process of the present manuscript, we become aware of a 
similar and related work, appeared as an e-print (arXiv:0908.1098), by \citet{cackett09}.

\bibliographystyle{aa}
\bibliography{refs}

\begin{thebibliography}{79}
\expandafter\ifx\csname natexlab\endcsname\relax\def\natexlab#1{#1}\fi

\bibitem[{{Agrawal} \& {Sreekumar}(2003)}]{agrawal03}
{Agrawal}, V.~K. \& {Sreekumar}, P. 2003, \mnras, 346, 933

\bibitem[{{Arnaud}(1996)}]{arnaud96}
{Arnaud}, K.~A. 1996, in Astronomical Society of the Pacific Conference Series,
  Vol. 101, Astronomical Data Analysis Software and Systems V, ed. G.~H.
  {Jacoby} \& J.~{Barnes}, 17--+

\bibitem[{{Ballantyne}(2004)}]{ballantyne04}
{Ballantyne}, D.~R. 2004, \mnras, 351, 57

\bibitem[{{Ballantyne} \& {Ross}(2002)}]{ballantyne02}
{Ballantyne}, D.~R. \& {Ross}, R.~R. 2002, \mnras, 332, 777

\bibitem[{{Ballantyne} {et~al.}(2001){Ballantyne}, {Ross}, \&
  {Fabian}}]{ballantyne01}
{Ballantyne}, D.~R., {Ross}, R.~R., \& {Fabian}, A.~C. 2001, \mnras, 327, 10

\bibitem[{{Ballantyne} \& {Strohmayer}(2004)}]{ballantyne04b}
{Ballantyne}, D.~R. \& {Strohmayer}, T.~E. 2004, \apjl, 602, L105

\bibitem[{{Barret} \& {Olive}(2002)}]{barret02}
{Barret}, D. \& {Olive}, J.-F. 2002, \apj, 576, 391

\bibitem[{{Barret} {et~al.}(2000){Barret}, {Olive}, {Boirin}, {Done},
  {Skinner}, \& {Grindlay}}]{barret00}
{Barret}, D., {Olive}, J.~F., {Boirin}, L., {et~al.} 2000, \apj, 533, 329

\bibitem[{{Barret} {et~al.}(2003){Barret}, {Olive}, \&
  {Oosterbroek}}]{barret03}
{Barret}, D., {Olive}, J.~F., \& {Oosterbroek}, T. 2003, \aap, 400, 643

\bibitem[{{Bhattacharyya} \& {Strohmayer}(2007)}]{bhattacharyya07}
{Bhattacharyya}, S. \& {Strohmayer}, T.~E. 2007, \apjl, 664, L103

\bibitem[{{Boirin} \& {Parmar}(2003)}]{boirin03}
{Boirin}, L. \& {Parmar}, A.~N. 2003, \aap, 407, 1079

\bibitem[{{Cabanac} {et~al.}(2009){Cabanac}, {Fender}, {Dunn}, \&
  {K{\"o}rding}}]{cabanac09}
{Cabanac}, C., {Fender}, R.~P., {Dunn}, R.~J.~H., \& {K{\"o}rding}, E.~G. 2009,
  \mnras, 396, 1415

\bibitem[{{Cackett} {et~al.}(2009){Cackett}, {Miller}, {Ballantyne}, {Barret},
  {Bhattacharyya}, {Boutelier}, {Miller}, {Strohmayer}, \&
  {Wijnands}}]{cackett09}
{Cackett}, E.~M., {Miller}, J.~M., {Ballantyne}, D.~R., {et~al.} 2009, ArXiv
  e-prints

\bibitem[{{Cackett} {et~al.}(2008){Cackett}, {Miller}, {Bhattacharyya},
  {Grindlay}, {Homan}, {van der Klis}, {Miller}, {Strohmayer}, \&
  {Wijnands}}]{cackett08}
{Cackett}, E.~M., {Miller}, J.~M., {Bhattacharyya}, S., {et~al.} 2008, \apj,
  674, 415

\bibitem[{{Church} \& {Baluci{\' n}ska-Church}(2001)}]{church01}
{Church}, M.~J. \& {Baluci{\' n}ska-Church}, M. 2001, \aap, 369, 915

\bibitem[{{Church} \& {Ba{\l}uci{\'n}ska-Church}(2004)}]{church04}
{Church}, M.~J. \& {Ba{\l}uci{\'n}ska-Church}, M. 2004, \mnras, 348, 955

\bibitem[{{D'A{\`i}} {et~al.}(2009){D'A{\`i}}, {Iaria}, {Di Salvo}, {Matt}, \&
  {Robba}}]{dai09}
{D'A{\`i}}, A., {Iaria}, R., {Di Salvo}, T., {Matt}, G., \& {Robba}, N.~R.
  2009, \apjl, 693, L1

\bibitem[{{D'A{\'{\i}}} {et~al.}(2007){D'A{\'{\i}}}, {{\.Z}ycki}, {Di Salvo},
  {Iaria}, {Lavagetto}, \& {Robba}}]{dai07}
{D'A{\'{\i}}}, A., {{\.Z}ycki}, P., {Di Salvo}, T., {et~al.} 2007, \apj, 667,
  411

\bibitem[{{D'Angelo} {et~al.}(2008){D'Angelo}, {Giannios}, {Dullemond}, \&
  {Spruit}}]{dangelo08}
{D'Angelo}, C., {Giannios}, D., {Dullemond}, C., \& {Spruit}, H. 2008, \aap,
  488, 441

\bibitem[{{den Herder} {et~al.}(2001){den Herder}, {Brinkman}, {Kahn},
  {Branduardi-Raymont}, {Thomsen}, {Aarts}, {Audard}, {Bixler}, {den Boggende},
  {Cottam}, {Decker}, {Dubbeldam}, {Erd}, {Goulooze}, {G{\"u}del}, {Guttridge},
  {Hailey}, {Janabi}, {Kaastra}, {de Korte}, {van Leeuwen}, {Mauche},
  {McCalden}, {Mewe}, {Naber}, {Paerels}, {Peterson}, {Rasmussen}, {Rees},
  {Sakelliou}, {Sako}, {Spodek}, {Stern}, {Tamura}, {Tandy}, {de Vries},
  {Welch}, \& {Zehnder}}]{denherder01}
{den Herder}, J.~W., {Brinkman}, A.~C., {Kahn}, S.~M., {et~al.} 2001, \aap,
  365, L7

\bibitem[{{Di Salvo} {et~al.}(2009){Di Salvo}, {D'A{\'{\i}}}, {Iaria},
  {Burderi}, {Dov{\v c}iak}, {Karas}, {Matt}, {Papitto}, {Piraino}, {Riggio},
  {Robba}, \& {Santangelo}}]{disalvo09}
{Di Salvo}, T., {D'A{\'{\i}}}, A., {Iaria}, R., {et~al.} 2009, \mnras, 398,
  2022

\bibitem[{{Di Salvo} {et~al.}(2002){Di Salvo}, {Farinelli}, {Burderi},
  {Frontera}, {Kuulkers}, {Masetti}, {Robba}, {Stella}, \& {van der
  Klis}}]{disalvo02}
{Di Salvo}, T., {Farinelli}, R., {Burderi}, L., {et~al.} 2002, \aap, 386, 535

\bibitem[{{Di Salvo} {et~al.}(2005){Di Salvo}, {Iaria}, {M{\'e}ndez},
  {Burderi}, {Lavagetto}, {Robba}, {Stella}, \& {van der Klis}}]{disalvo05}
{Di Salvo}, T., {Iaria}, R., {M{\'e}ndez}, M., {et~al.} 2005, \apjl, 623, L121

\bibitem[{{di Salvo} {et~al.}(2004){di Salvo}, {Santangelo}, \&
  {Segreto}}]{disalvo04}
{di Salvo}, T., {Santangelo}, A., \& {Segreto}, A. 2004, Nuclear Physics B
  Proceedings Supplements, 132, 446

\bibitem[{{Done} \& {Gierli{\'n}ski}(2006)}]{done06}
{Done}, C. \& {Gierli{\'n}ski}, M. 2006, \mnras, 367, 659

\bibitem[{{Done} {et~al.}(2002){Done}, {{\.Z}ycki}, \& {Smith}}]{done02}
{Done}, C., {{\.Z}ycki}, P.~T., \& {Smith}, D.~A. 2002, \mnras, 331, 453

\bibitem[{{Dullemond} \& {Spruit}(2005)}]{dullemond05}
{Dullemond}, C.~P. \& {Spruit}, H.~C. 2005, \aap, 434, 415

\bibitem[{{Fabian} {et~al.}(1989){Fabian}, {Rees}, {Stella}, \&
  {White}}]{fabian89}
{Fabian}, A.~C., {Rees}, M.~J., {Stella}, L., \& {White}, N.~E. 1989, \mnras,
  238, 729

\bibitem[{{Fiocchi} {et~al.}(2007){Fiocchi}, {Bazzano}, {Ubertini}, \&
  {Zdziarski}}]{fiocchi07}
{Fiocchi}, M., {Bazzano}, A., {Ubertini}, P., \& {Zdziarski}, A.~A. 2007, \apj,
  657, 448

\bibitem[{{Ford} {et~al.}(1998){Ford}, {van der Klis}, {van Paradijs},
  {M{\'e}ndez}, {Wijnands}, \& {Kaaret}}]{ford98}
{Ford}, E.~C., {van der Klis}, M., {van Paradijs}, J., {et~al.} 1998, \apjl,
  508, L155

\bibitem[{{Forman} {et~al.}(1978){Forman}, {Jones}, {Cominsky}, {Julien},
  {Murray}, {Peters}, {Tananbaum}, \& {Giacconi}}]{forman78}
{Forman}, W., {Jones}, C., {Cominsky}, L., {et~al.} 1978, \apjs, 38, 357

\bibitem[{{Galloway} {et~al.}(2008){Galloway}, {Muno}, {Hartman}, {Psaltis}, \&
  {Chakrabarty}}]{galloway08}
{Galloway}, D.~K., {Muno}, M.~P., {Hartman}, J.~M., {Psaltis}, D., \&
  {Chakrabarty}, D. 2008, \apjs, 179, 360

\bibitem[{{Gierli{\'n}ski} \& {Done}(2002)}]{gierlinski02}
{Gierli{\'n}ski}, M. \& {Done}, C. 2002, \mnras, 331, L47

\bibitem[{{Gierli{\'n}ski} {et~al.}(2008){Gierli{\'n}ski}, {Done}, \&
  {Page}}]{gierlinski08}
{Gierli{\'n}ski}, M., {Done}, C., \& {Page}, K. 2008, \mnras, 388, 753

\bibitem[{{Gilfanov} {et~al.}(2003){Gilfanov}, {Revnivtsev}, \&
  {Molkov}}]{gilfanov03}
{Gilfanov}, M., {Revnivtsev}, M., \& {Molkov}, S. 2003, \aap, 410, 217

\bibitem[{{Gladstone} {et~al.}(2007){Gladstone}, {Done}, \&
  {Gierli{\'n}ski}}]{gladstone07}
{Gladstone}, J., {Done}, C., \& {Gierli{\'n}ski}, M. 2007, \mnras, 378, 13

\bibitem[{{Gottwald} {et~al.}(1989){Gottwald}, {Haberl}, {Langmeier},
  {Hasinger}, {Lewin}, \& {van Paradijs}}]{gottwald89}
{Gottwald}, M., {Haberl}, F., {Langmeier}, A., {et~al.} 1989, \apj, 339, 1044

\bibitem[{{Haberl} \& {Titarchuk}(1995)}]{haberl95}
{Haberl}, F. \& {Titarchuk}, L. 1995, \aap, 299, 414

\bibitem[{{Hasinger} \& {van der Klis}(1989)}]{hasinger89}
{Hasinger}, G. \& {van der Klis}, M. 1989, \aap, 225, 79

\bibitem[{{Hiemstra} {et~al.}(2009){Hiemstra}, {Soleri}, {M{\'e}ndez},
  {Belloni}, {Mostafa}, \& {Wijnands}}]{hiemstra09}
{Hiemstra}, B., {Soleri}, P., {M{\'e}ndez}, M., {et~al.} 2009, \mnras, 394,
  2080

\bibitem[{{Homan} {et~al.}(2009){Homan}, {Kaplan}, {van den Berg}, \&
  {Young}}]{homan09}
{Homan}, J., {Kaplan}, D.~L., {van den Berg}, M., \& {Young}, A.~J. 2009, \apj,
  692, 73

\bibitem[{{Iaria} {et~al.}(2009){Iaria}, {D'A{\'{\i}}}, {di Salvo}, {Robba},
  {Riggio}, {Papitto}, \& {Burderi}}]{iaria09}
{Iaria}, R., {D'A{\'{\i}}}, A., {di Salvo}, T., {et~al.} 2009, \aap, 505, 1143

\bibitem[{{Inogamov} \& {Sunyaev}(1999)}]{inogamov99}
{Inogamov}, N.~A. \& {Sunyaev}, R.~A. 1999, Astronomy Letters, 25, 269

\bibitem[{{Jansen} {et~al.}(2001){Jansen}, {Lumb}, {Altieri}, {Clavel}, {Ehle},
  {Erd}, {Gabriel}, {Guainazzi}, {Gondoin}, {Much}, {Munoz}, {Santos},
  {Schartel}, {Texier}, \& {Vacanti}}]{jansen01}
{Jansen}, F., {Lumb}, D., {Altieri}, B., {et~al.} 2001, \aap, 365, L1

\bibitem[{{Langmeier} {et~al.}(1987){Langmeier}, {Sztajno}, {Hasinger},
  {Truemper}, \& {Gottwald}}]{langmeier87}
{Langmeier}, A., {Sztajno}, M., {Hasinger}, G., {Truemper}, J., \& {Gottwald},
  M. 1987, \apj, 323, 288

\bibitem[{{Lin} {et~al.}(2007){Lin}, {Remillard}, \& {Homan}}]{lin07}
{Lin}, D., {Remillard}, R.~A., \& {Homan}, J. 2007, \apj, 667, 1073

\bibitem[{{Maccarone} \& {Coppi}(2003)}]{maccarone03}
{Maccarone}, T.~J. \& {Coppi}, P.~S. 2003, \mnras, 338, 189

\bibitem[{{Makishima} {et~al.}(1986){Makishima}, {Maejima}, {Mitsuda}, {Bradt},
  {Remillard}, {Tuohy}, {Hoshi}, \& {Nakagawa}}]{makishima86}
{Makishima}, K., {Maejima}, Y., {Mitsuda}, K., {et~al.} 1986, \apj, 308, 635

\bibitem[{{Matt} {et~al.}(2003){Matt}, {Guainazzi}, \& {Maiolino}}]{matt03}
{Matt}, G., {Guainazzi}, M., \& {Maiolino}, R. 2003, \mnras, 342, 422

\bibitem[{{M{\'e}ndez} \& {Belloni}(2007)}]{mendez07}
{M{\'e}ndez}, M. \& {Belloni}, T. 2007, \mnras, 381, 790

\bibitem[{{Merloni} {et~al.}(2000){Merloni}, {Fabian}, \& {Ross}}]{merloni00}
{Merloni}, A., {Fabian}, A.~C., \& {Ross}, R.~R. 2000, \mnras, 313, 193

\bibitem[{{Meyer-Hofmeister} {et~al.}(2005{\natexlab{a}}){Meyer-Hofmeister},
  {Liu}, \& {Meyer}}]{meyer05}
{Meyer-Hofmeister}, E., {Liu}, B.~F., \& {Meyer}, F. 2005{\natexlab{a}}, \aap,
  432, 181

\bibitem[{{Meyer-Hofmeister} {et~al.}(2005{\natexlab{b}}){Meyer-Hofmeister},
  {Liu}, \& {Meyer}}]{meyer-hofmeister05}
{Meyer-Hofmeister}, E., {Liu}, B.~F., \& {Meyer}, F. 2005{\natexlab{b}}, \aap,
  432, 181

\bibitem[{{Miller} {et~al.}(2006{\natexlab{a}}){Miller}, {Homan}, \&
  {Miniutti}}]{miller06b}
{Miller}, J.~M., {Homan}, J., \& {Miniutti}, G. 2006{\natexlab{a}}, \apjl, 652,
  L113

\bibitem[{{Miller} {et~al.}(2006{\natexlab{b}}){Miller}, {Homan}, {Steeghs},
  {Rupen}, {Hunstead}, {Wijnands}, {Charles}, \& {Fabian}}]{miller06a}
{Miller}, J.~M., {Homan}, J., {Steeghs}, D., {et~al.} 2006{\natexlab{b}}, \apj,
  653, 525

\bibitem[{{Nayakshin} \& {Kazanas}(2002)}]{nayakshin02}
{Nayakshin}, S. \& {Kazanas}, D. 2002, \apj, 567, 85

\bibitem[{{Nayakshin} {et~al.}(2000){Nayakshin}, {Kazanas}, \&
  {Kallman}}]{nayakshin00}
{Nayakshin}, S., {Kazanas}, D., \& {Kallman}, T.~R. 2000, \apj, 537, 833

\bibitem[{{Olive} {et~al.}(2003){Olive}, {Barret}, \&
  {Gierli{\'n}ski}}]{olive03}
{Olive}, J.-F., {Barret}, D., \& {Gierli{\'n}ski}, M. 2003, \apj, 583, 416

\bibitem[{{Pandel} {et~al.}(2008){Pandel}, {Kaaret}, \& {Corbel}}]{pandel08}
{Pandel}, D., {Kaaret}, P., \& {Corbel}, S. 2008, \apj, 688, 1288

\bibitem[{{Papitto} {et~al.}(2009){Papitto}, {Di Salvo}, {D'A{\`i}}, {Iaria},
  {Burderi}, {Riggio}, {Menna}, \& {Robba}}]{papitto09}
{Papitto}, A., {Di Salvo}, T., {D'A{\`i}}, A., {et~al.} 2009, \aap, 493, L39

\bibitem[{{Piraino} {et~al.}(2007){Piraino}, {Santangelo}, {di Salvo},
  {Kaaret}, {Horns}, {Iaria}, \& {Burderi}}]{piraino07}
{Piraino}, S., {Santangelo}, A., {di Salvo}, T., {et~al.} 2007, \aap, 471, L17

\bibitem[{{Popham} \& {Sunyaev}(2001)}]{popham01}
{Popham}, R. \& {Sunyaev}, R. 2001, \apj, 547, 355

\bibitem[{{Raymond}(1993)}]{raymond93}
{Raymond}, J.~C. 1993, \apj, 412, 267

\bibitem[{{Reis} {et~al.}(2009){Reis}, {Fabian}, \& {Young}}]{reis09}
{Reis}, R.~C., {Fabian}, A.~C., \& {Young}, A.~J. 2009, ArXiv e-prints

\bibitem[{{Revnivtsev} \& {Gilfanov}(2006)}]{revnivtsev06}
{Revnivtsev}, M.~G. \& {Gilfanov}, M.~R. 2006, \aap, 453, 253

\bibitem[{{Ross} \& {Fabian}(2005)}]{ross05}
{Ross}, R.~R. \& {Fabian}, A.~C. 2005, \mnras, 358, 211

\bibitem[{{R{\'o}{\.z}a{\'n}ska} {et~al.}(2002){R{\'o}{\.z}a{\'n}ska},
  {Dumont}, {Czerny}, \& {Collin}}]{rozanska02}
{R{\'o}{\.z}a{\'n}ska}, A., {Dumont}, A., {Czerny}, B., \& {Collin}, S. 2002,
  \mnras, 332, 799

\bibitem[{{Rykoff} {et~al.}(2007){Rykoff}, {Miller}, {Steeghs}, \&
  {Torres}}]{rykoff07}
{Rykoff}, E.~S., {Miller}, J.~M., {Steeghs}, D., \& {Torres}, M.~A.~P. 2007,
  \apj, 666, 1129

\bibitem[{{Shimura} \& {Takahara}(1995)}]{shimura95}
{Shimura}, T. \& {Takahara}, F. 1995, \apj, 445, 780

\bibitem[{{Str{\"u}der} {et~al.}(2001){Str{\"u}der}, {Briel}, {Dennerl},
  {Hartmann}, {Kendziorra}, {Meidinger}, {Pfeffermann}, {Reppin}, {Aschenbach},
  {Bornemann}, {Br{\"a}uninger}, {Burkert}, {Elender}, {Freyberg}, {Haberl},
  {Hartner}, {Heuschmann}, {Hippmann}, {Kastelic}, {Kemmer}, {Kettenring},
  {Kink}, {Krause}, {M{\"u}ller}, {Oppitz}, {Pietsch}, {Popp}, {Predehl},
  {Read}, {Stephan}, {St{\"o}tter}, {Tr{\"u}mper}, {Holl}, {Kemmer}, {Soltau},
  {St{\"o}tter}, {Weber}, {Weichert}, {von Zanthier}, {Carathanassis}, {Lutz},
  {Richter}, {Solc}, {B{\"o}ttcher}, {Kuster}, {Staubert}, {Abbey}, {Holland},
  {Turner}, {Balasini}, {Bignami}, {La Palombara}, {Villa}, {Buttler},
  {Gianini}, {Lain{\'e}}, {Lumb}, \& {Dhez}}]{struder01}
{Str{\"u}der}, L., {Briel}, U., {Dennerl}, K., {et~al.} 2001, \aap, 365, L18

\bibitem[{{Taam} {et~al.}(2008){Taam}, {Liu}, {Meyer}, \&
  {Meyer-Hofmeister}}]{taam08}
{Taam}, R.~E., {Liu}, B.~F., {Meyer}, F., \& {Meyer-Hofmeister}, E. 2008, \apj,
  688, 527

\bibitem[{{Titarchuk}(1994)}]{titarchuk94}
{Titarchuk}, L. 1994, \apj, 434, 570

\bibitem[{{Turner} {et~al.}(2001){Turner}, {Abbey}, {Arnaud}, {Balasini},
  {Barbera}, {Belsole}, {Bennie}, {Bernard}, {Bignami}, {Boer}, {Briel},
  {Butler}, {Cara}, {Chabaud}, {Cole}, {Collura}, {Conte}, {Cros}, {Denby},
  {Dhez}, {Di Coco}, {Dowson}, {Ferrando}, {Ghizzardi}, {Gianotti}, {Goodall},
  {Gretton}, {Griffiths}, {Hainaut}, {Hochedez}, {Holland}, {Jourdain},
  {Kendziorra}, {Lagostina}, {Laine}, {La Palombara}, {Lortholary}, {Lumb},
  {Marty}, {Molendi}, {Pigot}, {Poindron}, {Pounds}, {Reeves}, {Reppin},
  {Rothenflug}, {Salvetat}, {Sauvageot}, {Schmitt}, {Sembay}, {Short},
  {Spragg}, {Stephen}, {Str{\"u}der}, {Tiengo}, {Trifoglio}, {Tr{\"u}mper},
  {Vercellone}, {Vigroux}, {Villa}, {Ward}, {Whitehead}, \& {Zonca}}]{turner01}
{Turner}, M.~J.~L., {Abbey}, A., {Arnaud}, M., {et~al.} 2001, \aap, 365, L27

\bibitem[{{van der Klis}(2004)}]{vanderklis04}
{van der Klis}, M. 2004, ArXiv Astrophysics e-prints

\bibitem[{{Verner} {et~al.}(1996){Verner}, {Ferland}, {Korista}, \&
  {Yakovlev}}]{verner96}
{Verner}, D.~A., {Ferland}, G.~J., {Korista}, K.~T., \& {Yakovlev}, D.~G. 1996,
  \apj, 465, 487

\bibitem[{{Wijnands} {et~al.}(1998){Wijnands}, {van der Klis}, {Mendez}, {van
  Paradijs}, {Lewin}, {Lamb}, {Vaughan}, \& {Kuulkers}}]{wijnands98}
{Wijnands}, R., {van der Klis}, M., {Mendez}, M., {et~al.} 1998, \apjl, 495,
  L39+

\bibitem[{{Wilms} {et~al.}(2000){Wilms}, {Allen}, \& {McCray}}]{wilms00}
{Wilms}, J., {Allen}, A., \& {McCray}, R. 2000, \apj, 542, 914

\bibitem[{{Zycki} \& {Czerny}(1994)}]{zycki94b}
{Zycki}, P.~T. \& {Czerny}, B. 1994, \mnras, 266, 653

\bibitem[{{Zycki} {et~al.}(1994){Zycki}, {Krolik}, {Zdziarski}, \&
  {Kallman}}]{zycki94}
{Zycki}, P.~T., {Krolik}, J.~H., {Zdziarski}, A.~A., \& {Kallman}, T.~R. 1994,
  \apj, 437, 597

\end{thebibliography}

\end{document}